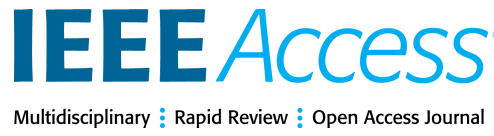

# Neuromorphic Vision Data Coding: Classifying and Reviewing the Literature

**CATARINA BRITES**[1], **(Member, IEEE), AND JOÃO ASCENSO**[2], **(Senior Member, IEEE)**

[1]Instituto de Telecomunicações, Instituto Universitário de Lisboa (ISCTE-IUL), 1649-026 Lisbon, Portugal

[2]Instituto de Telecomunicações, Instituto Superior Técnico, University of Lisbon, 1049-001 Lisbon, Portugal

Corresponding author: Catarina Brites (catarina.brites@lx.it.pt)

This work was supported in part by RayShaper SA, Valais, Switzerland, through the Project ''Event Aware Sensor Compression;'' and in part by the Fundação para a Ciência e a Tecnologia (FCT), Portugal, through the Project ''Deep Compression: Emerging Paradigm for Image Coding'' under Grant PTDC/EEI-COM/7775/2020.

**ABSTRACT** In recent years, visual sensors have been quickly improving towards mimicking the visual information acquisition process of human brain by responding to illumination changes as they occur in time rather than at fixed time intervals. In this context, the so-called *neuromorphic vision sensors* depart from the conventional frame-based image sensors by adopting a paradigm shift in the way visual information is acquired. This new way of visual information acquisition enables faster and asynchronous per-pixel responses/recordings driven by the scene dynamics with a very high dynamic range and low power consumption. However, depending on the application scenario, the emerging neuromorphic vision sensors may generate a large volume of data, thus critically demanding highly efficient coding solutions in order applications may take full advantage of these new, attractive sensors' capabilities. For this reason, considerable research efforts have been invested in recent years towards developing increasingly efficient neuromorphic vision data coding (NVDC) solutions. In this context, the main objective of this paper is to provide a comprehensive overview of NVDC solutions in the literature, guided by a novel classification taxonomy, which allows better organizing this emerging field. In this way, more solid conclusions can be drawn about the current NVDC *status quo*, thus allowing to better drive future research and standardization developments in this emerging technical area.

**INDEX TERMS** Dynamic vision sensor, event camera, neuromorphic vision data coding, spike camera, taxonomy.

## I. INTRODUCTION

Neuromorphic vision sensors are bio-inspired sensors that try to mimic the sensing behavior of a biological retina. These emerging sensors pose a paradigm shift in the visual information acquisition (sensing) model, the so-called *frameless paradigm*, where visual information is no longer acquired as a 2D matrix, i.e., *frame*. As it is well-known, in the conventional *frame-based paradigm*, all sensor pixels acquire visual information simultaneously, independently of the scene dynamics, at regular time intervals (i.e., constant framerate). However, in the emerging *frameless paradigm*, each sensor pixel is in charge of controlling its own visual information acquisition process, in an asynchronous and independent way, according to the scene dynamics, thus producing a variable data rate output; *scene dynamics* refers to the change of lightning/illumination conditions and/or motion in the scene and/or sensor/camera motion.

By incorporating ''intelligent'' pixels, frameless-based (i.e., neuromorphic) vision sensors provide interesting advantages over conventional frame-based image sensors, such as high temporal resolution (smaller time interval at which a sensor pixel can react/respond to the scene dynamics), very high dynamic range, low latency, and low power consumption. These are rather compelling properties notably in scenarios that are particularly challenging to the conventional frame-based image sensors, such as the ones involving visual scenes with high-speed motion and/or uncontrolled lighting conditions, where usually this type of (frame-based) image sensors fail to provide good performance; autonomous driving, drones and robotics are just a few examples of increasingly relevant applications in humans' daily lives where those challenging scenarios occur and, thus, can benefit from the neuromorphic vision sensors usage. Moreover, neuromorphic vision sensors may also have application in industrial automation, visual surveillance, augmented reality, and mobile environments, where fast response, high dynamic range or low power consumption are critically needed. Due to its potential and availability, neuromorphic vision sensors, also known in the literature as *event-based sensors*, *dynamic vision sensors*, *silicon retinas*, *spike cameras*, *asynchronous image sensors* or *frameless imaging sensors*, are nowadays attracting a great deal of attention by the research community from both academia and industry.

Currently, there are two main types of neuromorphic vision cameras, the so-called *event cameras* and *spike cameras*, which, roughly speaking, differ mainly on two aspects: 1) the way visual information is acquired, i.e., sampled, and consequently on the output data produced (*event data* versus *spike data*); and 2) the visual information they represent (moving areas only versus both static and moving areas of the visual scene acquired). The *event cameras* follow a *differential sampling model* in which time-domain changes in the incoming light intensity, i.e., *temporal contrast*, are pixel-wise detected and compared to a threshold, triggering a so-called *event* if it exceeds the threshold. On the other hand, the *spike cameras* follow an *integral sampling model* in which time-domain accumulation of the incoming light intensity is carried out pixel-wise and compared to a threshold, firing a so-called *spike* if the threshold is exceeded. In this context, *events* are triggered by event sensor pixels 'observing' scene's moving objects only, while *spikes* are fired by spike sensor pixels 'observing' both static and moving objects of the scene to be acquired. While the *dynamic vision sensor* (DVS) [1], [2], [3] was the first event camera to be made commercially available, among several nowadays available [4], the so-called *Vidar* camera [5], [6] is the only spike camera currently reported in the literature and, to the best of the authors knowledge, is not commercially available.

Differently from the *absolute intensity value* (simultaneously) outputted by every pixel in a conventional frame-based image sensor (grayscale/color value resulting from incoming light intensity accumulation over a specific exposure time), an *event* is typically represented by a 4-tuple $\langle x, y, t, p\rangle$ (see FIGURE 1). This ordinary representation of an event contains the *location* $(x, y)$ within the (event-based) sensor (pixel coordinates or addresses) where the event occurred, the *timestamp* $t$, i.e., the time at which the event was triggered, and the *polarity* $p$ of the event, indicating weather a light intensity increase ($p = 1$ or 'ON' event) or decrease ($p = -1$ or 'OFF' event) occurred; thus, $(x, y)$, $t$, and $p$ constitute the *event location information*, the *event time information*, and the *event polarity information*, respectively. This 4-tuple event representation is sometimes referred to in the literature as *address event representation* (*AER*), as this is the data representation used by the AER communication protocol to asynchronously transmit information (such as events) from the sensor to the event camera output or between asynchronous chips. As far as the *spike* is concerned, it is typically associated to an 'ON'- or 'OFF'- value, corresponding to a spike fired ('1') or not ('0'). Although a (spike-based) sensor pixel can fire spikes asynchronously and continuously (at an arbitrary time), the spike camera currently reported in the literature (the Vidar camera) only reads out the spikes fired, also known as *spike firing states*, periodically with a fixed time interval $T$ (in the order of $\mu$s) and does that for every sensor pixel. This means that, at sampling time $t = T$, the spike *firing state* ('0' or '1' value) of every pixel is read, forming a (binary) *spike frame*, with the height and width of the sensor pixel array; naturally, as the time passes, spike frames are formed one after the other, creating an 3D array of binary spike frames.

In the related literature, the event camera pixel output is most of the times called *event sequence* or *event stream* while the spike camera pixel output is typically known as *spike train*. The asynchronous *sensor event sequence/sensor spike train*, resulting from the (pixel) event sequences/spike trains of all sensor pixels, represent, therefore, the visual information 'observed' by the emerging event/spike sensors. It is worth noting that the sensor event sequence/sensor spike train can be further processed elsewhere in the camera or by a vision application for several purposes/tasks, including a more efficient representation (through coding). In this paper, *neuromorphic vision data* (NVD) is the generic terminology used to refer to both event and spike data.

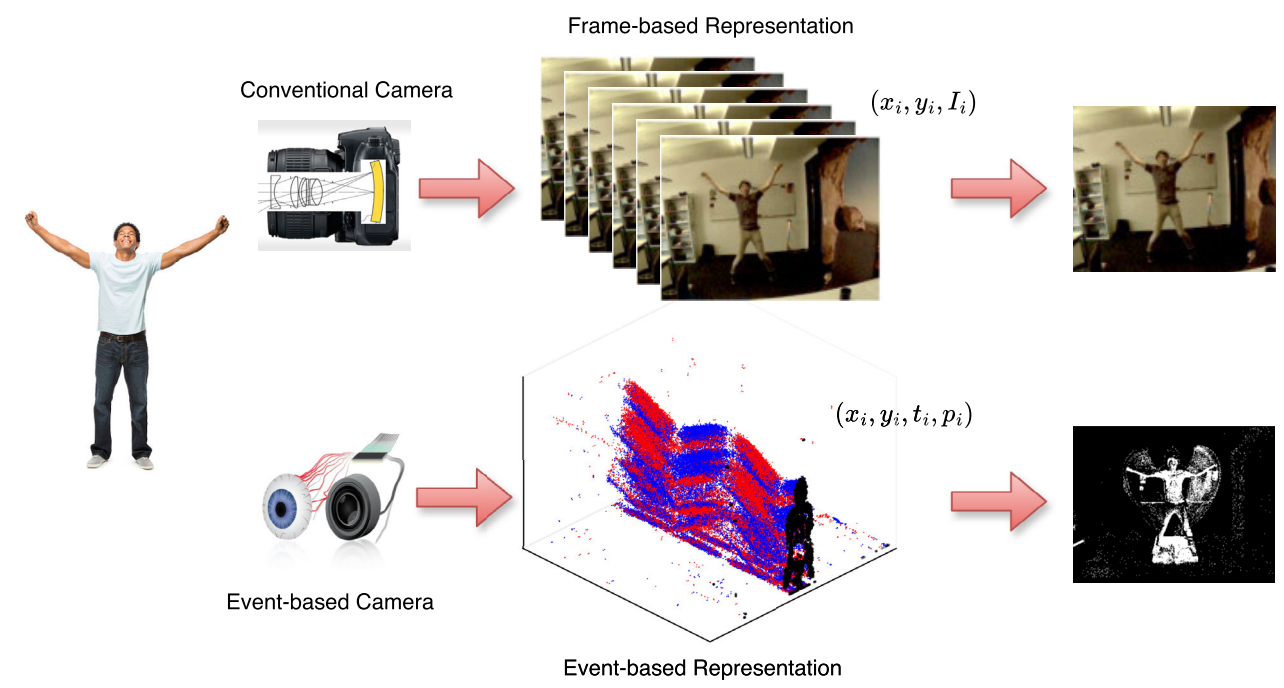

**FIGURE 1. Comparative illustration of (conventional) frame-based versus (emerging) event-based visual information representation.**

Despite the significant differences between neuromorphic vision data and conventional frame-based images, particularly in terms of the type and spatio-temporal characteristics of the visual information they represent, numerous algorithms have been recently developed to reconstruct images and videos from neuromorphic vision data, e.g., [7], [8]; this is largely because it allows existing image and video processing applications, typically designed to take image-based inputs, to use neuromorphic vision data. Either way, neuromorphic vision data are already successfully used in different tasks, including object tracking [9], [10], object recognition [11], [12], high-speed motion estimation [13], [14], HDR image reconstruction [15], [16], simultaneous localization and mapping (SLAM) [17], [18], among others.

Neuromorphic vision sensors achieve high temporal resolution, typically in the order of $\mu$s, which is equivalent to a high framerate (in the order of MHz) in conventional (frame-based) image sensors. Hence, depending on the application scenario and, consequently, on the scene dynamics complexity, the amount of raw data generated by the neuromorphic vision sensors may be substantial. The output data of neuromorphic vision sensors based on a differential sampling model (i.e., event cameras) tend to be sparse in the spatial dimension, as only pixels corresponding to moving/illumination changing areas trigger events, while exhibiting high temporal correlation. As for neuromorphic vision sensors based on an integral sampling model (i.e., spike cameras), the output data also exhibit high temporal correlation but are typically denser than the event camera output data in the spatial dimension, as pixels corresponding to both static and moving areas may fire spikes. Therefore, regardless of the sampling model, highly efficient neuromorphic visual data coding solutions are very much needed, especially considering the limited transmission and storage resources of the main applications of neuromorphic (particularly event-based) vision, such as embedded systems.

In the last few years, several (almost two tens) neuromorphic visual data coding (NVDC) solutions have been proposed, mainly driven by the growing availability of this type of cameras and the advantages they offer to scenarios where the conventional frame-based image cameras struggle to perform well; it is worth noting that, although some neuromorphic vision cameras may carry an inertial measurement unit (IMU), no coding solution for this type of data is reviewed in this paper as it has also not been found in the relevant NVDC literature. Nevertheless, a comparison of their performances is still a rather difficult task since the reported performance results have been obtained most of the times under different test conditions (even the evaluation methodologies are sometimes different) and there is no public software available to obtain comparable results. While some NVDC solutions target a real-time processing scenario, where low latency and low complexity in the encoding/decoding process are critical requirements, other NVDC solutions target scenarios of data storage or streaming for later processing, where complexity and latency requirements are relaxed in favor of high compression ratios. While the coding process in the former scenario may occur after data acquisition by the camera, in the latter scenarios the coding process can be performed by an application processor after some initial transmission. In any case, coding is always needed. Event rate can reach tens or hundreds of millions of events per second and, without coding, storage systems would quickly become overwhelmed and real-time streaming over bandwidth-constrained networks, such as wireless or IoT networks, would be impractical. Moreover, by reducing the data volume, coding also enables faster data transfer and processing, and reduces storage costs, especially for large datasets needed in AI applications or benchmarking, among other advantages.

Acknowledging the practical importance of developing efficient NVDC solutions, JPEG has recently launched an exploration activity on event-based vision, denominated JPEG XE [19]. The main goal of JPEG XE is to "*create and develop a standard to represent such events in an efficient way allowing interoperability between sensing, storage, and processing, targeting machine vision applications*" [19]. To achieve this goal, JPEG XE is currently focused on defining the use cases and requirements for potential standardization of the coding of events [20].

In this context, this paper first proposes a meaningful classification taxonomy for NVDC solutions that allows to identify and abstract their differences, commonalities, and relationships. Guided by this classification taxonomy, a large set of relevant NVDC solutions available in the literature is then reviewed; for a comprehensive overview of the NVDC field, both event-based and spike-based NDVC solutions will be considered. This paper does not purposely include any performance evaluation based on experimental results since its target is conceptual and algorithmic; in fact, it may still be early to derive final quantitative conclusions on the best NVDC coding approaches, considering the lack of technical maturity of most of these coding solutions, still requiring additional research. Therefore, the main objective of this paper is not to propose a novel NVDC solution, nor to provide a comparative performance evaluation of NVDC solutions, but rather to organize and classify a technical area that has received many contributions in recent years. This type of paper is essential to gather a systematic, high-level, and more abstract view of the field to further launch solid and consistent advancements in this emerging technical area.

With this purpose in mind, the rest of this paper is organized as follows: Section II will propose a classification taxonomy for the many NVDC solutions available in the literature, while Section III will provide an exhaustive review of the NVDC solutions available in the literature driven by the proposed taxonomy. Section IV will present an overview of the datasets used in the available NVDC literature, while Section V will present an overview of the performance evaluation metrics and relevant anchors used for benchmarking also in the available NVDC literature. Section VI will present some final remarks and, finally, Section VII will present some relevant challenges associated with NVDC, that emerged from the exhaustive literature reviewing.

## II. NVDC: PROPOSING A CLASSIFICATION TAXONOMY

The increasing popularity of neuromorphic vision sensors and, consequently, the increasing availability of neuromorphic vision content, has motivated the development of many coding solutions. Since multiple technical approaches have been adopted for the NVDC solutions available in the literature, it is essential to identify their main commonalities, differences, and relationships, thus providing a better understanding of the full neuromorphic vision data coding landscape and promising future research and standardization directions. In this context, this section first proposes a meaningful classification taxonomy and will after exercise it by referencing and classifying the many NVDC solutions found in the literature according to the proposed taxonomy; Section III will then go one step further by reviewing those solutions with a level of detail that allows to understand the involved key concepts and designs, although naturally not as detailed as the corresponding referenced papers. In the following sub-sections, the proposed classification dimensions for the taxonomy will be proposed first. After, the classes for each taxonomy classification dimension will be proposed. The classification dimensions and the classes within each dimension have been defined based on the exhaustive reviewing of (almost) two tens of NVDC solutions available in the literature in order a robust taxonomy could be defined [21], [22], [23], [24], [25], [26], [27], [28], [29], [30], [31], [32], [33], [34], [35], [36], [37], [38], [39]; this list of references is also an useful contribution of this paper.

### A. TAXONOMY CLASSIFICATION DIMENSIONS

This section presents and defines the classification dimensions for the taxonomy proposed for NVDC solutions. After an exhaustive study of the NVDC solutions available in the literature [21], [22], [23], [24], [25], [26], [27], [28], [29], [30], [31], [32], [33], [34], [35], [36], [37], [38], [39], it was concluded that the most appropriate taxonomy classification dimensions are:

1) ***Raw Data Type:*** Refers to the type of (raw) elementary information that is asynchronously outputted by each sensor pixel, e.g., an *event* or *spike*, and that is targeted by coding; the raw data type is intimately linked to the neuromorphic vision sensor type and its visual data acquisition model.
2) ***Fidelity:*** Refers to the fidelity with which the data are coded. Depending on the application domain, the data to be coded, hereafter referred to as *raw input data*, may correspond to the sensor output data or to the output data of an (optional) pre-processing module, placed in between the sensor and the (NVD) encoder. The pre-processing module may involve data filtering or data sampling, and might be used, for instance, to remove noise data from the sensor output data sequence or to control the amount of sensor output data to be effectively coded (i.e., fed to the NVD encoder). Since this (optional) pre-processing step takes place out of the encoder module, it is not directly related to the coding process and, thus, it is out of the scope of the proposed taxonomy (i.e., no taxonomy dimension is associated to it).
3) ***Data Structure:*** Refers to the way the raw input data sequence, i.e., sequence of elements of a specific raw data type at the NVD encoder input, is arranged/transformed to be then coded while exploiting the available spatial and temporal redundancies; depending on the adopted fidelity, this may involve data partitioning, temporal aggregation, data sampling, and data conversion.
4) ***Basic Coding Unit:*** Refers to the basic processing entity in which the structured data are further divided for coding purposes.
5) ***Components:*** Refers to the specific constituent, i.e., basic, elements of the raw data type in the structured data to be directly coded; depending on the raw data type, the components may be the basic elements of the ordinary raw data type representation (introduced in Section I), e.g., the *polarity*, or of an alternative representation with basic elements obtained from the ordinary raw data type representation during data structuring, e.g., the *time interval*.
6) ***Components Coding Approach:*** Refers to the way in which the components of the structured data are coded to reach a more compact neuromorphic visual data coded representation, e.g., independently or jointly.
7) ***Prediction:*** Refers to the way the component-wise spatial and temporal correlations in the basic processing entity of the structured data are exploited to create a lower energy signal, the so-called *residue*.
8) ***Transform:*** Refers to the way the remaining correlation in the component-wise residue signal is exploited to reach a more compact energy representation, usually in some type of frequency domain. A key issue related to the Transform dimension is that the impact of transform on the performance of NVDC solutions is a poorly understood domain (only 2 out of 19 coding solutions reviewed use transform); this might be a possible area of future research investment.
9) ***Quantization:*** Refers to the way the remaining correlation in the component-wise residue signal or the transform coefficients is exploited to create a lower energy signal, the so-called *quantized signal*. A key issue related to the Quantization dimension is that the impact of quantization on the performance of NVDC solutions is also a poorly understood domain (only 4 out of 19 coding solutions reviewed use quantization); this might be also a possible area of future research investment, notably if lossy coding is targeted, as quantization induces some loss or distortion in the decoded data.

Using these dimensions, each NVDC solution may be characterized by a taxonomy *classification path* connecting a set of classes along these dimensions, thus allowing to identify commonalities through the overlapping of the corresponding classification paths.

### B. CLASSES FOR EACH CLASSIFICATION DIMENSION

Using the proposed taxonomy classification dimensions, it is now necessary to define the classes within each classification dimension, naturally based on the NVDC solutions already available in the literature. After exhaustive analysis, the following classes are proposed for each dimension:

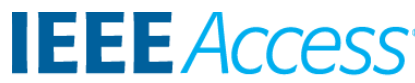

1) ***Raw Data Type*** – In terms of raw data type, the following classes are proposed:
   a) ***Event***: Asynchronous and independent response of a vision sensor pixel to a detected time-domain (incoming) light intensity relative change (i.e., above a preset threshold), so-called *temporal contrast*, at a precise time instant. It is typically represented by the 4-tuple $\langle x, y, t, p \rangle$ containing the *location* $(x, y)$ within the sensor (pixel coordinates) where the event was triggered, the *timestamp* $t$ at which the event was triggered, and the *polarity* $p$ of the event indicating weather a light intensity increase ($p = 1$) or decrease ($p = -1$) occurred.
   b) ***Spike***: Asynchronous and independent response of a vision sensor pixel to an accumulated time-domain (incoming) light intensity exceeding a preset threshold at a precise time instant. It is typically represented by the 3-tuple $\langle x, y, f \rangle$ containing the *location* $(x, y)$ within the sensor (pixel coordinates) where the spike was fired, and the *spike firing state* $f$ indicating whether a spike has been fired ($f = 1$) or not ($f = 0$). For the spike camera currently reported in the literature (the Vidar camera), the spike firing state of every pixel is read periodically, with a fixed time interval $T$, forming a (binary) spike frame; as the time passes by, spike frames are formed one after the other and the temporal information associated to the spike occurrences within each spike frame is inherently embedded in the respective spike frame index ($kT$). For this reason, no time information is explicitly included in the spike (3-tuple) representation.
2) ***Fidelity*** – In terms of fidelity, the following classes are proposed:
   a) ***Lossless***: Codecs keeping the original (i.e., raw input) data fidelity, meaning that the decoded and original data are mathematically equal (up to a certain precision, if required).
   b) ***Lossy***: Codecs not keeping the original (i.e., raw input) data fidelity, typically to increase the compression factor; high fidelity may still be achieved with the appropriate coding parameters configuration.
3) ***Data Structure*** – In terms of data structure (DS), the following classes are proposed:
   a) ***3D Point Set***: The raw input data sequence, corresponding to a sequence of elements of a specific raw data type at the NVD encoder input, are arranged as a set of points in the 3-dimensional (3D) space with the temporal dimension playing the role of a geometric dimension, notably the $Z$ coordinate axis; thus, each 3D point is defined with Cartesian coordinates $(x, y, z)$ and possibly an attribute, e.g., the polarity $p$. This data structure is typically spatially $(x, y)$ sparse and preserves the original information (i.e., number of elements and values) of all raw data type components, without involving any data temporal aggregation or sampling. The 3D Point Set data structure is suitable to be coded with standard-based point cloud geometry coding solutions, e.g., G-PCC [41], or with coding schemes involving point-based geometry processing.
   b) ***Cuboid Grid***: The raw input data sequence, corresponding to a sequence of elements of a specific raw data type at the NVD encoder input, are arranged in a space-time grid of *cuboids* (or cubes if all its 3 dimensions are equal); each *cuboid* represents a local spatio-temporal neighborhood of elements of a specific raw data type. While the grid resolution in the spatial dimensions ($X$ and $Y$) is typically regular, meaning that the cuboids' length on those dimensions is the same over the entire sensor resolution, in the temporal dimension the grid resolution may be regular or irregular, depending on the criterion used to split the sensor data on that dimension; for example, the temporal grid resolution can be determined from external information, e.g., the time interval between RGB images in a DAVIS-like camera, or by task-related motion requirements. This data structure tends to be spatially denser than the 3D Point Set data structure, as the distribution of the elements (events/spikes) in the spatial dimensions tends to concentrate into small areas (such as a cuboid spatial area) corresponding, for instance, to objects' movement; naturally, the elements distribution varies with the scene dynamics. The Cuboid Grid data structure also preserves the original information (i.e., number of elements and values) of all raw data type components, without involving any data temporal aggregation or sampling. This data structure is suitable to be coded with solutions based on spatio-temporal volumes such as cuboids; this type of data structure allows exploiting spatial correlation between spatially neighboring (i.e., in the $X$ and $Y$ dimensions) cuboids and the temporal correlation within a cuboid and between co-located (in the spatial dimensions) temporally adjacent cuboids.
   c) ***1D Array of Elements***: The raw input data sequence, corresponding to a sequence of elements of a specific raw data type at the NVD encoder input, are arranged as a 1-dimensional (1D) array, i.e., as a single sequence, of elements that may or may not be ordered in some component(s), e.g., elements may be ordered by the triggering/firing time. This data structure is dense, in the sense that each array position refers to a triggering/firing output, and preserves the original information (i.e., number of elements and values) of all raw data type components without involving any data temporal aggregation or sampling. The 1D Array of Elements data structure is suitable to be coded with element-wise coding solutions or coding schemes that exploit components' correlation through prediction or entropy coding strategies within small portions of the 1D array, i.e., chunks, with fixed or arbitrary size.

d) ***3D Array of Frames***: The raw input data sequence, corresponding to a sequence of elements of a specific raw data type at the NVD encoder input, are converted into a 3-dimensional array of *frames*, e.g., by pixel-wise polarity accumulation or pixel-wise counting the elements of a specific raw data type (e.g., events) over a given time interval or by simply sampling elements over time; a *frame* is basically a 2D array, with the sensor spatial resolution, whose entry (pixel) values are typically obtained from event temporal aggregation at each pixel location (e.g., a pixel may correspond to an accumulated polarity value or a histogram count). The 3D Array of Frames data structure is typically denser (than the 3D Point Set data structure) and usually does not preserve the original information (i.e., number of elements and values) of some raw data type components, as it typically involves data temporal aggregation and polarity accumulation. This data structure is suitable to be coded with standard video coding solutions, e.g., HEVC, or with coding solutions inspired on the intra and inter coding modes adopted in the standard image/video coding solutions or even with lookup table (LUT)-based coding schemes; this type of data structure allows exploiting spatial correlation (within a frame) and/or correlation between temporally adjacent frames in the 3D array.

4) ***Basic Coding Unit*** – In terms of basic coding unit (BCU), the following classes are proposed:

a) ***Single Element***: Basic processing entity of structured data corresponding to an individual element of a given raw data type. This basic coding unit is characteristic of element-by-element processing methods and allows spatio-temporal correlation exploitation, although it might not be that efficient due to the small support region for correlation exploitation (1 element only).

b) ***Chunk***: Basic processing entity of structured data corresponding to a group of $N_C$ elements of a given raw data type, i.e., a *chunk*. The chunk size $N_C$ may be fixed or adjusted dynamically using, for instance, a criterion associated to the triggering time instant. This basic coding unit may potentiate high spatio-temporal correlation exploitation depending on the criterion used to define the elements belonging to a chunk (and the chunk size), which impacts on the similarities between neighboring elements within and between chunks.

c) ***Polyhedron***: Basic processing entity of structured data corresponding to a 3D shape with flat polygonal faces, straight edges, and sharp vertices, containing elements of a given raw data. A cuboid, i.e., a $N_X \times N_Y \times N_T$ (3D) volume of elements, is a particular case of this basic coding unit, which corresponds to a polyhedron with six quadrilateral (flat) faces; $N_X$ and $N_Y$ stand for the volume length in the spatial dimensions $X$ and $Y$, respectively, and $N_T$ stands for the volume length in the temporal dimension. In a cuboid, $N_X$ and $N_Y$ are usually set to the same value, which in theory can vary between 1 (corresponding to a sensor pixel) and the sensor height and width, respectively, while $N_T$ is typically adjusted dynamically using a criterion associated to possible (task-related) motion requirements. Depending on the scene dynamics, other 3D shape polyhedrons (than cuboids) may allow a finer adaptation to the motion characteristics, by aggregating in it neighboring elements (of a given raw data) with similar motion characteristics; in this case, a motion plane-based representation of the polyhedron, characterized by a more or less complex set of parameters (e.g., represented as a set of the 3D shape's vertices and edges), can be for instance adopted. This basic coding unit is suitable for a fine adaptation to the motion characteristics and allows exploring the spatial or spatio-temporal correlation between spatially neighboring polyhedrons and the temporal correlation within or between co-located (in the spatial dimensions) temporally adjacent 3D polyhedrons.

d) ***Group of Frames***: Basic processing entity of structured data corresponding to a set, i.e., a group, of $N_I$ frames, typically contiguous in time, where a frame typically corresponds to a 2D array ($N_X \times N_Y$) of values of a specific component (e.g., polarity), accumulated over a certain time interval $\Delta$ or sampled over the temporal dimension; $N_X$ and $N_Y$ correspond to the frame height and width, which are typically equal to the sensor height and width, respectively. A single frame is a particular case of this basic coding unit, which corresponds to $N_I$ equal to one. In case $N_I = 1$, this basic coding unit may allow exploiting spatial correlation (i.e., within a frame) with standard-based image/video coding solutions, e.g., HEVC Intra. When $N_I > 1$, this basic coding unit may potentiate the spatio-temporal correlation exploitation (within and between frames of a group of frames), facilitating the adoption of coding schemes according to the components to be coded, e.g., different coding schemes for location and polarity.

5) ***Components*** – In terms of components, the following classes are proposed:

   a) ***Location***: Information to be coded includes location data, i.e., 2D coordinates ($x$, $y$) identifying the position in the sensor where an asynchronous sensor output of a specific raw data type was triggered; this information is typically present in the representation of both *event* and *spike* raw data types.
   b) ***Timestamp***: Information to be coded includes time data, i.e., temporal information identifying when an asynchronous sensor output of a specific raw data type was triggered at a given pixel location; this information is present in the *event* raw data type representation only.
   c) ***Polarity***: Information to be coded includes polarity data, i.e., 1-bit code indicating the variation sign of the temporal contrast (corresponding to a light intensity increase or decrease) associated to an asynchronous sensor output of a specific raw data type triggered at a given pixel location; this information is present in the *event* raw data type representation only.
   d) ***Time Interval***: Information to be coded includes time interval data, i.e., information indicating the time interval between two consecutive occurrences of asynchronous outputs of the sensor triggered at a given pixel location; this information is present in the *spike* and *event* raw data type representation. Contrary to the previous component classes, the Time Interval class corresponds to information that is not directly recorded by the sensor; the time interval information is converted from the sequence of spike firing states ('1'/'0' values) or timestamps outputted by each sensor pixel along time. It is worth noting that, for the spike raw data type, the time interval data, known as *inter-spike interval* (*ISI*) data, are typically characterized by higher spatio-temporal correlation than the (recorded) spike firing state data, which makes them preferable for coding purposes [23]; this is, in fact, the reason why (all) the spike-based NVDC solutions available in the literature code the time interval data (i.e., ISI data) instead of the data recorded by the sensor (spike firing states).

6) ***Components Coding Approach*** – In terms of components coding approach, the following classes are proposed:

   a) ***Independent***: The components data, associated to a specific raw data type representation, are coded separately, meaning that each component is independently encoded/decoded from the remaining one(s); this may involve using different coding strategies to code each component, possibly with knowledge on some coding information of other component, e.g., its elements' coding order. Depending on the adopted data structure, component embedding in the data structure itself or on the BCU may be involved; this means that the embedded components are not directly coded.
   b) ***Joint***: The components data, associated to a specific raw data type representation, are coded jointly, meaning that all the components are encoded/decoded jointly; this typically involves using a single coding strategy to code all the components at once. Depending on the adopted data structure, component embedding in the data structure itself or on the BCU may be involved; this means that the embedded components are not directly coded.

7) ***Prediction*** – In terms of prediction, the following classes are proposed:

   a) ***None***: No prediction is applied at all.
   b) ***Intra***: The basic processing unit of the structured data is coded while exploiting the correlation among its elements only; this is called *Intra* prediction since the correlation is exploited within a single basic processing unit. This prediction class may involve exploitation of temporal correlation only or both spatial and temporal correlation, typically depending on whether the basic processing unit spans over temporal dimension only (corresponding to a sequence of elements of a given pixel) or over both spatial and temporal dimensions (corresponding to sequences of elements of pixels in a spatial neighborhood). It may also involve the definition of different Intra coding modes, which are adaptively selected for different parts of the content depending on the spatial distribution of the elements.
   c) ***Inter***: The basic processing unit of the structured data is coded while exploiting the correlation between basic processing units in some spatio-temporal neighborhood, considering motion; this is called *Inter* prediction as, taking motion into account, previous encoded basic processing units may be used as reference to the current basic processing unit coding. This prediction class may involve exploitation of temporal correlation only (in this case between spatially co-located neighboring basic processing units) or both spatial and temporal correlation depending on whether motion in a spatio-temporal neighborhood is considered or not. It may also involve the definition of different Inter coding modes, which are adaptively selected for different parts of the content depending on the spatio-temporal distribution of the elements.

d) ***Hybrid***: The basic processing unit of the structured data is coded while exploiting the correlation among its elements and between elements belonging to basic processing units in some spatio-temporal neighborhood; this is called *hybrid* prediction and may involve the definition of (Intra and Inter) coding modes, which are adaptively selected for different parts of the content.

8) ***Transform*** – In terms of transform, the following classes are proposed:

a) ***None***: No transform is applied at all.

b) ***Block-based***: A transform is applied to some appropriate signal or residual signal, structured as a regular block; this includes for example the discrete cosine transform (DCT). The transform may be fixed or hand-crafted (e.g., DCT), adaptive (e.g., Karhunen–Loève Transform - KLT), or learned (e.g., deep learning-based).

9) ***Quantization*** – In terms of quantization, the following classes are proposed:

a) ***None***: No quantization is applied at all.

b) ***Uniform***: A quantization is applied to some appropriate signal or residual signal where the quantization levels are uniformly spaced.

c) ***Non-Uniform***: A quantization is applied to some appropriate signal or residual signal where the quantization levels are unequally spaced.

An overview of the proposed classification taxonomy is shown in FIGURE II; note that the arrows simply intend to highlight example connection paths between classes along the nine dimensions. Because the Raw Data Type dimension is a strong dividing factor, the next section will provide a comprehensive review of the NVDC solutions available in the literature guided by the proposed taxonomy's Raw Data Type dimension, to better understand the involved key concepts and designs in NVDC, both within and between classes of the Raw Data Type dimension.

### C. NVDC LITERATURE OVERVIEW

To experience and appreciate the power of the proposed classification taxonomy, this section offers a summary table where the NVDC references available in the literature are classified according to the proposed taxonomy, see TABLE 1; this table allows identifying related NVDC solutions with respect to one or more taxonomy dimensions. Since the references are chronologically ordered, it is easy to see the NVDC technical approaches evolution along time. For example, it is interesting to note that event is the most adopted Raw Data Type class while lossy is the most adopted Fidelity class.

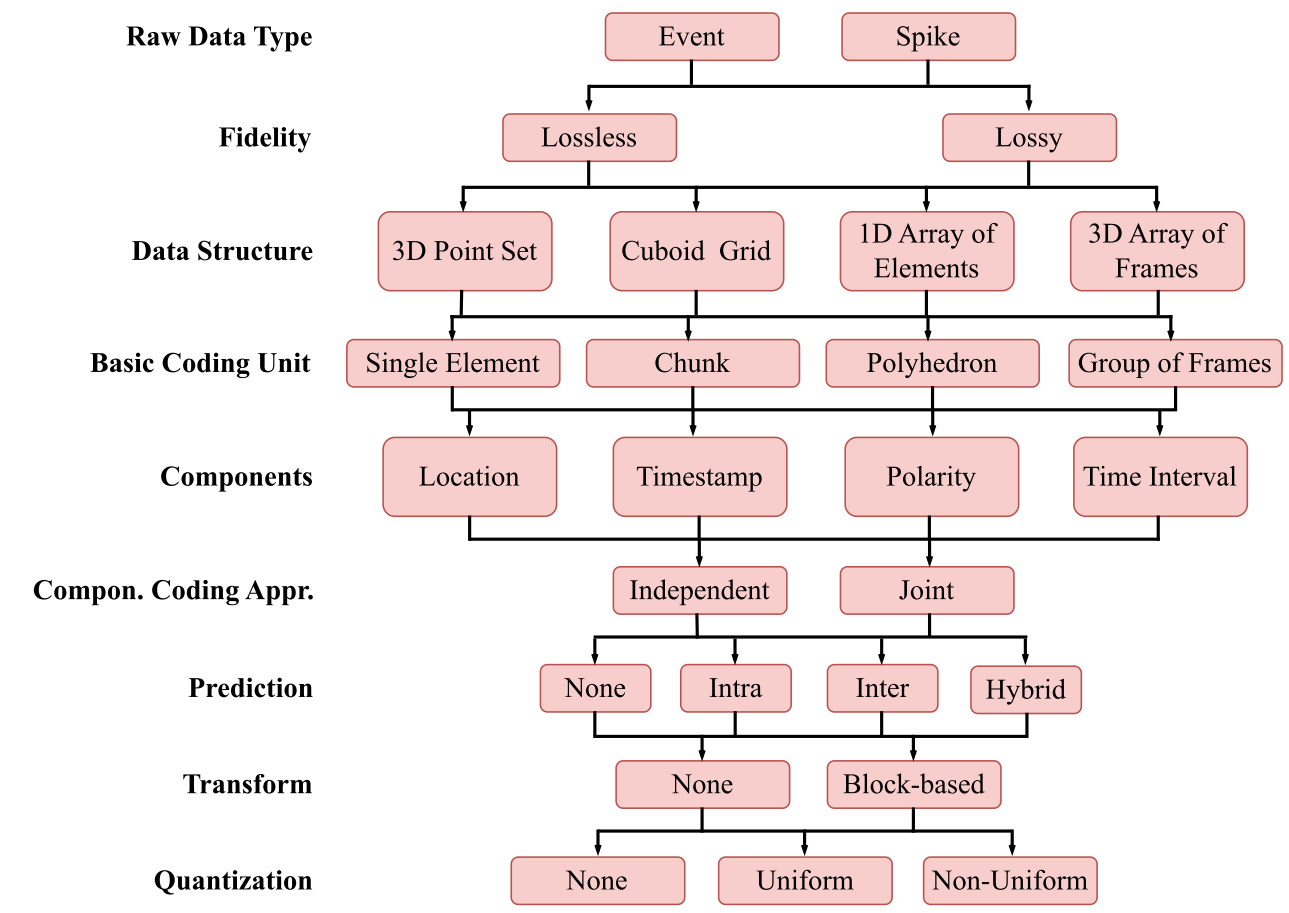

**FIGURE 2. Overview of the proposed NVDC classification taxonomy.**

Moreover, the first learning-based NVDC solution has just emerged in 2022, adopting a lossy coding approach.

For each entry in TABLE 1, whenever there is only one class in the 'Pred.', 'Transf.' and/or 'Quant.' columns for more than one class in the 'Compon.' column, this means that all components listed in 'Compon.' are predicted, transformed and/or quantized in the same way; for each entry in TABLE 1, whenever there is more than one class in the 'Compon.'/'Pred.'/'Quant.' dimensions, they are separated by '&'. In TABLE 1, '/' is used to identify different variants proposed for a specific classification dimension in each reference. The symbol '?' in TABLE 1 means that not enough information is provided in the reference to clarify the respective classification.

From TABLE 1, the following conclusions can be drawn:

- The vast majority of the NVDC solutions currently available in the literature, including the most recent ones, are event-based (17 out of 19). This seems a natural consequence of the significantly higher number of event-based cameras commercially available [4] when compared to the single spike-based camera currently reported in the literature (the Vidar camera [5]).
- Most of the NVDC solutions are lossy (12 out of 19), although some of those solutions (e.g., [25], [28], [30], [34], [37] and [38]) are classified as lossy not because they employ lossy coding tools but rather because they perform lossy operations, such as temporal quantization and/or polarity accumulation, during the data structuring of the raw input data sequence; while temporal quantization induces some precision loss in the timestamp component of the raw input data sequence, polarity accumulation induces some precision loss in the polarity component of the raw input data sequence. While lossless NVDC seems to be more appropriate for typical NVDC use cases, lossy coding with adjustable CR/quality tradeoff may also be adequate/advantageous for certain application scenarios, e.g., on-demand slow motion, for further reducing the events coding data rate (compared to lossless coding); please refer to [20] for more details on possible event-based vision use cases and requirements under study on the recent JPEG exploration activity on event-based vision (JPEG XE [19]). Moreover, it is worth noting that lossy coding has always been targeted in spike-based NVDC solutions.

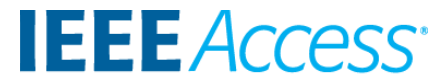

**TABLE 1.** **NVDC solutions in the literature classified according to the proposed taxonomy.**

| Ref. | Raw Data Type | Fidelity | Data Structure | Basic Coding Unit | Compon. | Compon. Coding Appr. | Pred. | Transf. | Quant. |
|---|---|---|---|---|---|---|---|---|---|
| Z. Bi *et al.*, 2018 [21] | Event | Lossless | Cuboid Grid | Polyhedron $[N_X \times N_Y \times N_T]$ | Location &<br>Polarity &<br>Timestamp | Independent | None/Intra-cuboid &<br>None &<br>Intra-cuboid | None | None |
| S. Dong *et al.*, 2019 [22] | Event | Lossless | Cuboid Grid | Polyhedron $[N_X \times N_Y \times N_T]$ | Location &<br>Polarity &<br>Timestamp | Independent | Hybrid-cuboid &<br>None &<br>Hybrid-cuboid | None | None |
| S. Dong *et al.*, 2019 [23] | Spike | Lossy | Cuboid Grid | Polyhedron $[1 \times 1 \times N_T]$ | Time Interval ‡ | Independent | Hybrid-cuboid | None | Non-Uniform |
| Y. Fu *et al.*, 2019 [24] | Event | Lossy | Cuboid Grid | Polyhedron $[N_X \times N_Y \times N_T]$ | Location &<br>Polarity &<br>Timestamp | Independent | Hybrid-cuboid &<br>None &<br>Hybrid-cuboid | None | Uniform?<br>&<br>None? &<br>Uniform? |
| N. Khan *et al.*, 2021 [25] | Event | Lossy | 3D Array of Frames † | Group of Frames $[N_I = 1]$ | Location (histogram) ‡ | Independent | Hybrid-frame | None | None |
| L. Zhu *et al.*, 2021 [26] | Spike | Lossy | Cuboid Grid | Polyhedron | Time Interval ‡ | Independent | Hybrid-polyhedron | Block-based | Non-Uniform |
| S. Banerjee *et al.*, 2021 [27] | Event | Lossy | Cuboid Grid † | Polyhedron $[N_X \times N_Y \times N_T]$ | Location &<br>Polarity ‡ | Independent | Intra-cuboid &<br>None | None | None |
| I. Schiopu and R. C. Bilcu, 2022 [28] | Event | Lossy/Lossless | 3D Array of Frames † | Group of Frames $[N_I = 8]$ | Location &<br>(accumulated) Polarity ‡ | Independent | None | None | None |
| A. Hasssan *et al.*, 2022 [29] | Event | Lossy | 3D Array of Frames † | Group of Frames $[N_I = 1]$ | Location &<br>Polarity ‡ | Joint | None | None | Uniform |
| I. Schiopu and R. C. Bilcu, 2022 [30] | Event | Lossy/Lossless | 3D Array of Frames † | Group of Frames | SAFE codec:<br>Location &<br>Time Interval ‡<br>MER codec:<br>(accumulated) Polarity ‡ | Independent | None | None | None |
| M. Martini *et al.*, 2022 [31] | Event | Lossless | 3D Point Set | Polyhedron $[N_X \times N_Y \times N_T]$ | Location &<br>Timestamp ‡ | Joint | None | None | None |
| I. Schiopu and R. C. Bilcu, 2022 [32] | Event | Lossless | 1D Array of Elements | Chunk | Location &<br>Polarity ‡ | Independent | Intra-chunk &<br>None | None | None |
| C. Wang *et al.*, 2023 [33] | Event | Lossless | 1D Array of Elements | Single Element | Location &<br>Polarity &<br>Timestamp | Independent | None &<br>None &<br>Intra-chunk | None | None |
| I. Schiopu and R. C. Bilcu, 2023 [34] | Event | Lossy/Lossless | 3D Array of Frames † | Group of Frames $[N_I = 1]$ | (accumulated) Polarity ‡ | Independent | None | None | None |

**TABLE 1.** ***(Continued.)*** **NVDC solutions in the literature classified according to the proposed taxonomy.**

| I. Schiopu and R. C. Bilcu, 2023 [35] | Event | Lossless | 1D Array of Elements | Chunk | Location & Polarity ‡ | Independent | Intra-chunk & None | None | None |
|---|---|---|---|---|---|---|---|---|---|
| B. Huang and T. Ebrahimi, 2023 [36] | Event | Lossless | 3D Point Set | Polyhedron [$N_X \times N_Y \times N_T$] | Location & Timestamp ‡ | Joint | None | None | None |
| S. Khaidem *et al.*, 2023 [37] | Event | Lossy | 3D Array of Frames † | Group of Frames [$N_I = 1$] | Location (histogram) ‡ | Independent | None | None | None |
| J. Adhuran *et al.*, 2024 [38] | Event | Lossy | 3D Point Set | Polyhedron [$N_X \times N_Y \times N_T$] | Location & Timestamp ‡ | Joint | None | Block-based | None |
| D. Stumpp *et al.*, 2024 [39] | Event | Lossy | 1D Array of Elements † | Single Element | Location & Polarity & Timestamp | Joint | None | None | None |

† In the data structure creation process, some lossy operations, such as timestamp quantization, polarity accumulation and/or event sampling, may be involved (for more details please refer to Section III).
‡ Components not directly coded are typically embedded in the data structure itself and/or in the basic coding unit processing order (for more details please refer to Section III).

- In the event-based NVDC solutions, there is a trend towards reducing the number of event's components to be directly coded by embedding one (or more) component(s), e.g., timestamp or polarity, in the way the events are arranged to be coded (i.e., in the data structure); a similar trend is observed in the spike-based NVDC solutions, as the location and temporal information are somehow embedded in the BCU processing order and in the spike firing states readout time by the spike camera, respectively. The embedding approach seems to allow some event/spike data rate saving in comparison to coding all the components of the event/spike representation.
- Some of the most recent event-based NVDC solutions jointly encode the event's components, in opposition to the independent coding approach followed by the former NVDC solutions. While the joint approach may allow some event data rate saving (in comparison to the components independent coding), it may also increase the coding solution complexity, as it becomes a multi-variate coding approach.
- In terms of the data structure/basic coding unit dimensions, cuboid grid/polyhedron (notably cuboid-like polyhedron) seem to have been the trend in the earlier NVDC solutions, but for the more recent ones (published since 2022) the trend is not that clear. A possible reason for this may be the willingness of exploiting, in the neuromorphic vision data context, new visual data coding strategies that meanwhile emerged (e.g., G-PCC and deep learning-based coding), and that, to be effective, require the adoption of more appropriate data structures/basic coding units (such as 3D point set/polyhedron and 3D array of frames/group of frames).
- Finally, regarding the coding tools addressed by the proposed taxonomy, it is possible to conclude from TABLE 1 that while (some sort of) prediction is used by most of the NVDC solutions available (10 out of 19), the same does not apply to transform, which is only used in 2 out 19 NVDC solutions (1 event-based solution and 1 spike-based solution). In terms of quantization, it is interesting to notice that, from the 4 NVDC solutions where the quantization tool is employed, only 2 of them are event-based and, from these 2, both adopt a uniform quantization; regarding the spike-based NVDC solutions, both use the quantization tool, with a non-uniform spacing of the quantization levels. Moreover, from the small set of NVDC solutions adopting the transform and quantization coding tools, it is possible to infer that the impact of those coding tools on the NVDC solutions' performance is a poorly understood domain; these might be two possible areas of future research.

## III. NVDC: REVIEWING GUIDED BY THE TAXONOMY'S RAW DATA TYPE DIMENSION

The NVDC area has received many contributions in the recent years and is considered critical for the future of visual data coding solutions. In the following, an exhaustive list of the neuromorphic vision data coding (NVDC) solutions available in the literature is presented together with a description of each solution; the solutions' comprehensive review is guided by the proposed taxonomy's Raw Data Type dimension and follows the chronological order within each class of the Raw Data Type dimension. Some performance results are also reported for each NVDC solution to better understand its strengths and weaknesses. However, it is important to stress that the performance results reported in sections III-A and III-B are, in most cases, not directly comparable between different coding solutions, as they have been obtained under different test conditions, e.g., different evaluating event/spike sequences sets (even when the same dataset is used) and/or different duration of the coded sequences. Sometimes even different evaluation methodologies are used in the coding solutions performance assessment; while some solutions evaluate their performances with respect to the raw input data, other solutions may evaluate its performance with respect to aggregated data obtained from the raw input data structuring process, typically involving operations that lead to a loss of information in the raw input data. An earlier performance comparison of lossless coding strategies, including the NVDC solution in [21] and some generic lossless data compression strategies adapted to NVDC, can be found in [42]. For an overview on the datasets and performance evaluation metrics/benchmarks adopted by the NVDC solutions reviewed in this section, please refer to sections IV and V, respectively.

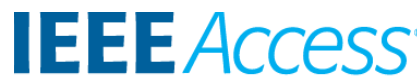

### A. EVENT-BASED NVDC SCHEMES

#### 1) SPIKE CODING FOR DYNAMIC VISION SENSORS [21]

In 2018, Bi et al. proposed a cube-based coding framework to losslessly code data generated from a DVS event camera (consisting of location, polarity, and timestamp) [21]. Although [21] uses the term *spike data* to refer to the DVS camera output data, in most NVDC literature (notably the most recent one) *event data* is the term commonly used to refer to the output data produced by an event camera, such as the DVS event camera (see Section I); for this reason, the term *event* has been adopted in this solution description.

In the proposed coding framework, which was the first made available to code event data, the *sensor event sequence* (outputted by the sensor pixel array) is first organized as a set of points in the 3D (space-time) volume, with the pixel location $(x, y)$ and the timestamp $t$ defining the 3D coordinate axes $X$, $Y$ and $Z$, respectively, and the polarity being the value attributed to each 3D point. An adaptive *macro-cube* partitioning of the sensor event sequence in the temporal (i.e., $Z$) dimension is then performed based on a binary-tree structure, targeting to obtain approximately the same number of events within each macro-cube; a *macro-cube* is a cube of event data in the 3D space whose size in $X$ and $Y$ dimensions corresponds to the sensor pixel array full spatial resolution and the size in $Z$ (temporal dimension) results from binary-tree partitioning. Each macro-cube is further split into $32 \times 32$ *event-cubes* along the spatial dimensions $(X, Y)$, constituting the basic unit for encoding and, thus, to exploit the (events) spatio-temporal redundancies. It is worth recalling that, in the related literature, *events' temporal redundancy* is related to the similarity between time intervals between consecutive events at a given pixel, typically resulting from a constant changing rate of luminance intensity (such as linear increase or decrease). *Events' spatial redundancy* is related to the similarity between events triggered by adjacent (sensor) pixels, typically resulting from the fact that adjacent pixels tend to simultaneously receive almost the same luminance intensities.

The proposed event-cube encoding procedure consists of separate encoding of (event) location, timestamp, and polarity data. The event location encoding process involves evaluating two intra-cube prediction modes designed to tackle different event spatial distributions, the so-called *address-prior mode* and *time-prior mode*, and selecting the one leading to the lowest rate cost; while the address-prior mode is designed for spatially sparse cubes, resulting from events scattering over the entire spatial resolution, the time-prior mode is designed for spatially dense cubes, resulting from a high events concentration over neighboring pixels. As intra-cube prediction modes, the address-prior mode and the time-prior mode only exploit the events correlation within an event-cube.

Thus, in the address-prior mode, the events within the cube to be encoded are first accumulated and stored in the so-called *location histogram*, a 2D array where each entry represents the number of events that were triggered in the corresponding pixel location within the cube; the location histogram is complemented with a *location histogram binary map*, a 2D array where each entry indicates whether events were triggered or not in the corresponding pixel location. Then, the location histogram and the location histogram binary map are (separately) fed into a context-based adaptive arithmetic entropy encoder, generating the *address-prior mode-based location coding bitstream*. For each pixel within the cube, the events' timestamps are differentially encoded with respect to the previous (event) timestamp, followed by context-based adaptive arithmetic entropy encoding, which generates the *address-prior mode-based timestamp coding bitstream.*

In the time-prior mode, it is first determined the so-called *center* point, i.e., the event pixel location within the cube that minimizes the spatial distance (in $x$ and $y$ pixel coordinates) to all the other event pixel locations within the cube to be encoded. Next, it is computed the displacement $(\Delta x, \Delta y)$ between the $(x, y)$ coordinates of every event (within the cube) and the center point coordinates $(x_c, y_c)$ previously determined; the set of $(\Delta x, \Delta y)$ displacements computed is then fed into a context-based adaptive arithmetic entropy encoder, generating the *time-prior mode-based location coding bitstream*. The events' timestamps (within the cube) are also differentially encoded with respect to the previous (event) timestamp (the coding order of the event timestamp follows the event location coding order), followed by context-based adaptive arithmetic entropy encoding, which generates the *time-prior mode-based timestamp coding bitstream*. The elementary encoded bitstreams associated to the (event) location and timestamp coding result then from the corresponding bitstreams generated by the intra-cube prediction mode with the lowest rate (considering both the location and timestamp rates).

As far as the event polarity coding is concerned, it is context-based adaptive arithmetic entropy encoded (with a coding order following the event location coding order of the intra-cube prediction mode with the lowest rate), generating the *polarity coding bitstream*. The elementary encoded bitstreams resulting from (event) location, timestamp and polarity coding are then multiplexed to generate the *event coding bitstream*.

Experimental results on the PKU-DVS event dataset, proposed in [21] for the event data coding algorithm evaluation, show an average compression ratio (over the whole dataset) of 19.52 with respect to the *raw event data size* (where each event is represented by 64 bits); hereafter, the compression ratio with respect to the raw event data size will be simply referred to as *compression ratio* (*CR*). Compared to the LZ77 and LZMA benchmarks, two Lempel-Ziv-based (generic) lossless coding algorithms, the proposed coding framework achieves an average CR 4.43× and 1.54× higher, respectively.

#### 2) SPIKE CODING FOR DYNAMIC VISION SENSOR IN INTELLIGENT DRIVING [22]

In 2019, Dong et al. extended the lossless event data coding solution in [21], by proposing an adaptive octree-based partitioning of the sensor event sequence into the so-called *coding tree cubes* in both spatial (*XY* axes) and temporal (*Z* axis) dimensions, to code event data generated from a DAVIS event camera [22]; DAVIS (Dynamic and Active Pixel Vision Sensor) [43] is an *hybrid sensor* that concurrently outputs event data (through a DVS sensor) and conventional intensity images/frames (through an active pixel sensor – APS).

In the proposed coding framework, each 64 × 64 × 32768 coding tree cube can be further adaptively divided along the spatial and temporal dimensions into smaller cubes, called *coding cubes*, using an octree structure, targeting to obtain approximately the same number of events within each smaller cube (i.e., coding cube); the coding cube constitutes, thus, the basic coding unit of the proposed coding solution.

The proposed coding cube encoding procedure consists of separate encoding of (event) location, timestamp, and polarity data. Similarly to [21], two intra-cube prediction modes, i.e., the address-prior mode and the time-prior mode, are evaluated and the one leading to the lowest rate cost is selected as the best prediction mode. However, in [22], the evaluation of each intra-cube prediction mode includes not only the (event) location and timestamp data coding but also the polarity data coding; the (event) location, timestamp, and polarity coding strategies adopted in [22] are similar to the ones described in the previous solution ([21]). Thus, the intra-cube prediction mode with the lowest sum of (event) location, timestamp, and polarity rates is selected as the best prediction mode; the elementary encoded bitstreams resulting from the location, timestamp and polarity coding for the best prediction mode are then multiplexed to generate the *event coding bitstream*.

Experimental results on the DDD17 dataset (only event data considered) show that the proposed solution slightly outperforms the lossless event coding solution in [21], achieving an average CR (with respect to the raw event data size) of 2.65 whereas the event coding solution in [21] achieves an average CR of 2.64. Compared to the LZ77 (Lempel-Ziv compression algorithm) and LZMA (Lempel-Ziv-Markov chain algorithm) lossless benchmarks, the proposed coding framework achieves an average CR 1.78× and 1.24× higher, respectively. An inter-cube prediction strategy, in which coding cubes previously encoded can be used as reference to predict the current coding cube, was also proposed in [22] but the coding performance achieved (average CR of 2.64) was rather similar to the one attained using only the intra-cube prediction strategy (average CR of 2.65).

#### 3) SPIKE CODING: TOWARDS LOSSY COMPRESSION FOR DYNAMIC VISION SENSOR [24]

Also in 2019, Fu et al. proposed the first lossy coding scheme to code event data generated from a DVS event camera [24]. The proposed solution extends the lossless coding framework in [22] to the lossy scenario by incorporating quantization of the prediction residual data together with an optimized inter-cube prediction and a new intra-cube prediction mode. It is worth noting that the details on the proposed coding scheme in [24] are scarce due to the paper length (1-page paper).

Experimental results on the MNIST-DVS dataset show the evolution of the average distortion and classification accuracy with the CR, for the proposed coding solution only; no details are given on how the CR, average distortion and classification accuracy were computed.

#### 4) TIME-AGGREGATION-BASED LOSSLESS VIDEO ENCODING FOR NEUROMORPHIC VISION SENSOR DATA [25]

In 2021, Khan et al. proposed the so-called *Time-Aggregation-based Lossless Video Encoding* (*TALVEN*) solution, a coding solution based on event aggregation in the temporal dimension and conventional lossless video coding, to code pseudo video sequences created from event data generated from a DAVIS event camera (consisting of location, polarity, and timestamp) [25].

The TALVEN solution starts by aggregating (i.e., accumulating), over a fixed time interval called *aggregation time interval*, the number of events triggered at each pixel location according to their polarity, creating two *polarity- based event frames*; a polarity-based event frame is, thus, a location histogram, i.e., a 2D array with the full sensor pixel array spatial resolution representing, at each pixel location, the event count for a given polarity in a fixed aggregation time interval.

This process of event aggregation over a fixed time interval, performed during the raw input data structuring, involves (raw input) event timestamps quantization, i.e., a lossy operation where all the (continuous) raw input (event) timestamps within the aggregation time interval are mapped to a single timestamp value, typically with reduced bit representation; event timestamps quantization induces, thus, some precision loss in the timestamp component of the raw input event sequence. Therefore, from the event coding pipeline point of view, which includes structuring of the raw input data towards subsequent coding (see Section II-A), the TALVEN solution is a lossy coding scheme; the (raw input) event timestamps quantization is, however, the only operation inducing loss of information in the TALVEN solution.

The two polarity-based event frames created from event aggregation are then concatenated side by side, creating a (new) bigger frame called *superframe*, targeting to take the most benefit of the conventional video coding techniques and, thus, achieving higher compression gains. This (event accumulation and polarity-based event frames concatenation) process is repeated over the whole duration of the sensor event sequence (i.e., raw input event sequence) and the resulting set of superframes, structured into a 3D array of superframes, is then treated as a *pseudo video sequence*; the pseudo video frames are not exactly conventional video frames, notably in terms of its content (value stored in each pixel is not an intensity value), hence the *pseudo* term. The pseudo video sequence is then HEVC losslessly encoded, generating the *event coding bitstream*; it is worth noting that, in the TALVEN solution, the polarity information is embedded within the video frame (each superframe concatenates both polarity-based event frames side by side) while the quantized timestamp information (for all the events in each video frame) is embedded in the frame number field of the encoded video sequence.

Experimental results on 10 (indoor and outdoor) event sequences of the DAVIS 240C dataset (only event data considered) show higher coding performance for medium to high aggregation time intervals compared to the event coding solution in [22], achieving CRs up to 4× higher. Besides the usual CR, between the raw event data size (in bits) and the event coding bitstream size (in bits), [25] also reports results on the CR between the raw (pseudo) video sequence size (in bits) and the event coding bitstream (in bits), called *video encoder CR*; the coding performance of the proposed solution in terms of video encoder CR follows a similar trend to the usual CR.

#### 5) LOSSY EVENT COMPRESSION BASED ON IMAGE-DERIVED QUAD TREES AND POISSON DISK SAMPLING [27]

Also in 2021, Banerjee et al. proposed the so-called *Poisson Disk Sampling-Lossy Event Compression* (*PDS-LEC*), a lossy coding solution based on Poisson disk sampling and quad-tree segmentation of intensity images to code event data generated from a DAVIS event camera (consisting of location, polarity, and timestamp) [27]; please recall that the DAVIS sensor [43] is an *hybrid sensor* that produces both event data (through a DVS sensor) and conventional intensity images (through an APS sensor). While the input of the proposed coding framework includes both event data (location, polarity, and timestamp) and RGB images, the work in [27] is focused on the event data coding only; RGB images are assumed to be coded in (some of) the experimental results but there is no reference on the coding solution employed to compress them.

In the PDS-LEC solution, a quad-tree structure is first derived for the intensity frame at time instant $t$, $I_t$, through dynamic programming (Viterbi algorithm), based on the decoded frame at time instant $t$-1, $\hat{I}_{t-1}$, targeting to guide the coding process of the event data triggered between time instants $t$-1 and $t$. Next, events triggered between time instants $t$-1 and $t$ are aggregated according to their polarity and temporally quantized into 16 bins, creating (two) *polarity-based event cuboid grids*; each (polarity-based) *event cuboid* is a volume of event data in a 3D space-time neighborhood and constitutes the basic coding unit of the PDS-LEC solution.

The (polarity-based) event cuboids are then adaptively sampled via Poisson disk sampling according to the priority established by the quad-tree based segmentation map; the priority of a 2D spatial region within the 3D space-time volume (of event data that have been triggered between time instants $t$-1 and $t$) is inversely proportional to the corresponding block size in the quad-tree based segmentation map. The higher the region priority, the higher the importance of keeping the respective events after sampling. Afterwards, the locations $(x, y)$ of the sampled events are differentially encoded followed by Huffman encoding, generating the *location coding bitstream*, while the polarity information of the sampled events is run-length encoded followed by Huffman encoding, generating the *polarity coding bitstream*. The elementary encoded bitstreams resulting from (event) location and polarity coding are then multiplexed to generate the *event coding bitstream*.

Experimental results on the DAVIS 240C dataset show higher compression performance, measured in terms of CR versus aggregation time interval, compared to the lossless and lossy event coding benchmark solutions in [22] and [25], respectively, achieving CRs up to 6× higher; it is not clear from [27], however, the precise conditions in which this lossy-lossless comparison was performed. More recently, the PDS-LEC solution has been adopted by the same authors for object tracking in a communication system called *host-chip architecture* [40]; in the proposed host-chip architecture, the data acquisition on the chip (event and intensity images) is driven by information transmitted by the host, an object tracking application, through a feedback channel. Experimental results show the robustness of the proposed system when compared to benchmark object tracking methods using DVS.

#### 6) LOSSLESS COMPRESSION OF EVENT CAMERA FRAMES [28]

In 2022, Schiopu and Bilcu proposed a performance-oriented, context-based image coding solution to code groups of event frames generated from a DVS-like camera (consisting of location, polarity, and timestamp), where the event location information and the event polarity information are encoded separately using different strategies [28].

In the proposed coding framework, the polarity of the events that occurred at each (sensor) pixel location are first accumulated (summed up) over a fixed time interval $\Delta$ and the (polarity) sum's sign $\{-1, 0, 1\}$ is then stored in the corresponding pixel location in a 2D array known as *event frame* (*EF*); an EF has, thus, the full sensor pixel array spatial resolution and represents, at each pixel location, the polarity of an event that represents all the events triggered in the accumulation time interval. Depending on the $\Delta$ value, the process of event aggregation over a fixed time interval $\Delta$ with polarity accumulation, performed during the raw input data structuring, may involve (raw input) event timestamps quantization and polarity accumulation, two operations that induce some precision loss in the timestamp and polarity components of the raw input event sequence, respectively; while in event timestamps quantization all the raw input (event) timestamps within the aggregation time interval are mapped to a single timestamp value, typically with reduced bit representation, in polarity accumulation, all the raw input (event) polarities within the aggregation time interval are summed up and converted to a single (representative) polarity value. In this context, from the event coding pipeline point of view, which includes structuring of the raw input data towards subsequent coding (see Section II-A), the proposed framework [28] is a lossy/lossless coding scheme; the (raw input) event timestamps quantization and polarity accumulation are the two operations inducing loss of information in the proposed coding framework.

The proposed process of event aggregation over a fixed time interval with polarity accumulation is repeated over the whole sensor event sequence duration and the resulting set of (synchronous) EFs is then structured into a 3D array of EFs, with each group of 8 (consecutive) EFs constituting the basic coding unit of the proposed coding framework.

For coding purposes, each group of 8 (consecutive) EFs is then represented by a pair of an *event map image* (*EMI*), storing the spatial information, and a *concatenated polarity vector* (*CPV*), storing the polarity information. The EMI is further represented by: i) a (2D) binary map (BM), signaling the pixel locations $(x, y)$ where at least one event has occurred in an EF; ii) the number of events per signaled position in BM; and iii) the EF indices, indicating the positions in the EF group of the EFs associated to the events previously identified.

The BM is encoded in raster scan order using template context modelling (TCM), where the context is computed using the causal neighborhood of the pixel to be encoded. The number of events is encoded bitplane by bitplane, starting with the least significant one in a 3-bitplane representation (maximum number of events is 8), also using template context modelling; the context for each bitplane is computed using the causal neighborhood of the current bitplane and a template context from the previously coded bitplane(s). The EF indices are encoded using adaptive Markov modelling (AMM). As far as the coding of polarity information is concerned, both AMM and TCM are applied, and the final encoding strategy is the one with the lowest estimated codelength. In the proposed coding solution, the *event coding bitstream* results from multiplexing the elementary bitstreams resulting from EMI and CPV encoding.

Experimental results on the DSEC dataset show that the proposed solution achieves CRs up to 5.8 (with respect to the raw event data size) for the time interval $\Delta = 10^{-6}$s. Compared to the conventional lossless video/image coding solutions used as benchmarks, HEVC (High Efficiency Video Coding), VVC (Versatile Video Coding), CALIC (Context-based Adaptive Lossless Image Coding) and FLIF (Free Lossless Image Format), [28] reports, for $\Delta = 10^{-6}$s, CR improvements of 198.01%, 238.94%, 125.04%, and 84.92%, respectively. Note that, when the 3D array of EFs is generated from the raw input (sensor) event sequence considering the time interval $\Delta = 10^{-6}$s (i.e., framerate of $10^6$ frames per second (fps)), all events of the raw input event sequence having the same timestamp are aggregated in one EF; this means that, for $\Delta = 10^{-6}$s, there is no (raw input) event timestamps quantization nor polarity accumulation, i.e., it corresponds to a lossless scenario. For the other time intervals $\Delta$ (i.e., framerates) evaluated, notably 5.555ms (180 fps), 1ms ($10^3$fps) and 0.1ms ($10^4$fps), [28] reports average coding performance improvements of 70.68%, 58.06% and 20.66% compared to HEVC, VVC and FLIF, respectively. This coding performance is measured in terms of a so-called *aggregation CR* (see Section V), i.e., a CR relative to the size (in bits) of an event (temporal) aggregation based raw input data structure, in this case the 3D array of EFs (where each EF pixel is represented by 2 bits, as 3 symbols are possible $\{-1, 0, 1\}$); please recall that the 3D array of EFs results from a process of event aggregation over a fixed time interval $\Delta$ with polarity accumulation (as described in the beginning of this section). It is worth noting that the input to the conventional image/video coding solutions are (event) images obtained by combining, through some mathematical function, the information enclosed in each set of 5 consecutive EFs (represented by 8-bit values), for improved coding performance of those codecs (please refer to [28] for more details).

#### 7) SPATIAL-TEMPORAL DATA COMPRESSION OF DYNAMIC VISION SENSOR OUTPUT WITH HIGH PIXEL-LEVEL SALIENCY USING LOW-PRECISION SPARSE AUTOENCODER [29]

Also in 2022, Hasssan et al. proposed the first learning-based lossy coding framework to code event frames created from event data generated from DVS-like cameras (consisting of location, polarity, and timestamp) [29].

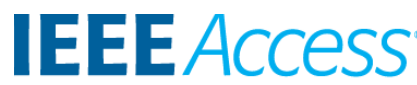

In the proposed framework, the sensor event sequence (i.e., raw input event sequence) is first converted into (synchronous) binary frames by sampling the (raw input) events $\langle x, y, p, t\rangle$ over the temporal dimension (timestamp $t$); however, this conversion process is not described in detail in [29]. The binary frames are then fed as input to the proposed low-precision sparse convolutional autoencoder architecture, where the encoder comprises 2 sparse convolutional layers (each including a ReLU activation function) and 2 max pooling layers, and the decoder includes 2 upsampling layers and 2 convolutional layers (each of which including also a ReLU activation function). To reduce the amount of computation and storage resources needed, low-precision 2-bit and 4-bit convolution operations were implemented during the training phase; full precision weights and activations of each convolution layer were passed to a quantization module before performing convolution operations, thus compressing the full precision weights and activations to 2-bit and 4-bit precision levels, respectively. The sparse compressed representation model (also known as *latent space*) is obtained by adding a L1 norm sparsity penalty term to the loss function during the training phase; the latent space corresponds to the *event coding bitstream* of the proposed coding solution.

Autoencoder reconstructed images are then used for inference by a classification network and an object detection network, both of which were independently trained on the original, i.e., raw input, (DVS) event data. It is worth noting that, contrary to the all the previously reviewed solutions, the learning-based lossy coding framework proposed in [29] was not evaluated standalone but in the context of specific computer vision tasks, notably event-based classification and object detection tasks. In the classification task, experimental results on the datasets MNIST-DVS (generated by converting the standard frame-based MNIST dataset to events), DvsGesture and Gen1 N-CARS show average CRs up to 29.1 with an accuracy drop of 3.0%. In the object detection task, experimental results on the Gen1 Automotive Detection dataset show an average CR of 11.9 with a drop of 0.07 in mean average precision.

#### 8) LOW-COMPLEXITY LOSSLESS CODING FOR MEMORY-EFFICIENT REPRESENTATION OF EVENT CAMERA FRAMES [30]

Still in 2022, Schiopu and Bilcu proposed a low complexity coding framework based on run-length and Elias coding to code, in a memory efficient way, event frames created from event data generated from a DVS-like camera (consisting of location, polarity, and timestamp) while targeting to be suitable for hardware implementation in low-cost event signal processing chips [30].

In the proposed coding framework, the polarity of the events triggered at each pixel location are first summed up over a fixed time interval $\Delta$ and the polarity sum's sign $\{-1, 0, 1\}$ is then stored in the corresponding pixel location in the so-called *event frame* (*EF*), as in [28]. Hence, similarly to solution 6) reviewed above ( [28]), the coding framework proposed in [30] is classified as lossy/lossless, since, depending on the $\Delta$ value, it may involve raw input event timestamps quantization and polarity accumulation, two operations that lead to some precision loss in the timestamp and the polarity components, respectively; these are, however, the only two operations inducing loss of information in the proposed coding framework.

The proposed process of event aggregation over a fixed time interval with polarity accumulation is repeated over the whole sensor event sequence duration (or any pre-defined time length) and the resulting (synchronous) EFs are then grouped together forming the so-called *EF volume* (i.e., a 3D array of EFs). The EF volume serves as input for the two proposed coding solutions: i) SAFE (*Simple And Fast lossless Event frame*) codec, for fast coding of large sets of EFs (thousands of EFs); and ii) MER (*Memory-Efficient Representation*) codec, for a memory-efficient representation of EFs while providing random access (RA) to any group of pixels within the EF volume. The EF volume constitutes, thus, the basic coding unit of the proposed coding solutions.

In the SAFE solution, the EF volume is further represented by a (2D) binary map (BM), signaling the pixel locations $(x, y)$ where at least one event was triggered in time, and a vector of concatenated time intervals (VCT), associated to each pixel location signaled in BM. Next, the run-length encoding scheme is adapted for coding both vectorized BM and VCT, by counting the number of consecutive event/no-event symbols in each, and then the Elias coding algorithm is applied to code those set of counts, generating the *event coding bitstream*.

In the MER solution, the EF volume (3D array of EFs) is further divided into a set of $8 \times 8 \times 8$ cubes, constituting the RA units, which are then arranged as a set of vectors V. As for the SAFE codec, the MER codec adapts the run-length coding scheme and Elias coding to code the set of vectors V, generating the *RA event coding bitstream*.

Experimental results on the DSEC dataset show that the proposed MER solution achieves an average aggregation CR (over the DSEC dataset) of 8.77 for $\Delta = 1$ms ($10^3$fps), with an *average EF encoding speed* of 9.07 ms per EF (ms/EF); please recall that the aggregation CR is a CR with respect to the size (in bits) of the (uncompressed) 3D array of EFs (see Section V), where each EF pixel is represented by 2 bits (as 3 symbols are possible $\{-1, 0, 1\}$). The so-called *average EF encoding speed* metric measures the (average) time needed to encode one EF (see TABLE 4). Experimental results also show that the MER solution provides fast and low complexity RA to any $8 \times 8 \times 8$ volume of pixels. For the same time interval $\Delta = 1$ms, the proposed SAFE solution achieves an average aggregation CR of 9.77, with an average EF encoding speed of 5.59 ms/EF. Compared to the conventional lossless video/image coding solutions used as benchmarks, HEVC, VVC and CALIC, [30] reports, for the best performing proposed solution (SAFE), average aggregation CR improvements of 34.57%, 24.62%, 35.51%, and 84.92%, respectively, with average EF encoding speed reductions of 92.79%, 99.97%, and 65.21%, respectively. For the same time interval $\Delta = 1$ms, the conventional lossless image coding solution FLIF outperforms the best performing proposed solution (SAFE) with an average aggregation CR 5% higher but requires a higher average EF encoding speed (26.10× higher). When the EF volume is generated from the sensor event sequence considering a time interval $\Delta = 10^{-6}$s ($10^6$fps), an average CR of 4073 is reported for the proposed SAFE solution, with an average EF encoding speed of 59.86 ms/EF; similarly to the solution 6) reviewed above ([28]), there is no (raw input) event timestamps quantization nor polarity accumulation for $\Delta = 10^{-6}$s (lossless scenario). For $\Delta = 10^{-6}$s, the proposed SAFE solution achieves average CR gains (over the DSEC dataset) 2.37×, 2.70×, 1.47×, and 1.76× higher than the ones obtained with HEVC, VVC, FLIF, and CALIC, respectively, while the average EF encoding speeds are 0.77×, 0.07×, 0.64×, and 5× the ones required by HEVC, VVC, FLIF, and CALIC, respectively. Please recall that, as in [28], the input to the conventional image/video coding solutions are (event) images obtained by combining, through some mathematical function, the information enclosed in each set of 5 consecutive EFs (represented by 8-bit values), for improved coding performance of those codecs (please refer to [28] for more details).

#### 9) LOSSLESS COMPRESSION OF NEUROMORPHIC VISION SENSOR DATA BASED ON POINT CLOUD REPRESENTATION [31]

In 2022, Martini et al. proposed a lossless coding solution based on standard point cloud compression to code event data generated from a DAVIS event camera (consisting of location, polarity, and timestamp) [31].

In the proposed solution, the sensor event sequence is organized as a set of points in 3D (space-time) volume, where the pixel location $(x, y)$ and the timestamp $t$ correspond to the 3D coordinate axes $X$, $Y$ and $Z$, respectively, and the polarity is the value attributed to each 3D point, thus resembling a 3D point cloud representation. By splitting the events (of the sensor event sequence) according to their polarity, two *3D point clouds* are obtained, one for each polarity.

Then, the standard Geometry-based Point Cloud Compression (G-PCC) codec [41] is applied to separately encode the geometry information, i.e., the $(x, y, t)$ triplet, of each 3D (polarity-based) point cloud. G-PCC applies a transformation to the input $(x, y, t)$ coordinates and structures the resulting data into voxelized octrees; the geometry is then losslessly coded with an octree of appropriate depth. The two elementary encoded bitstreams resulting from applying G-PCC to each 3D (polarity-based) point cloud are then multiplexed to generate the *event coding bitstream*.

Experimental results on the DAVIS 240C dataset (only event data considered) show higher compression performance, measured in terms of CR, when compared to the lossless benchmark solution in [21] and LZMA, a Lempel-Ziv-based (generic) lossless coding algorithm; CRs up to 30% and 49.4% higher are reported in [31] with respect to [21] and LZMA, respectively.

#### 10) LOW-COMPLEXITY LOSSLESS CODING OF ASYNCHRONOUS EVENT SEQUENCES FOR LOW-POWER CHIP INTEGRATION [32]

Also in 2022, Schiopu and Bilcu proposed the so-called *Low-complexity Lossless Compression of AsynchRonous Event Sequence* (*LLC-ARES*), a lossless coding framework based on a novel event data representation, called *same-timestamp* representation, and triple threshold-based range partition (TTP) algorithm. The LLC-ARES framework codes event data generated from a DVS-like camera (consisting of location, polarity, and timestamp) while targeting to be suitable for hardware implementation in low power event signal processing chips [32].

In the proposed LLC-ARES framework, the sensor event sequence is first divided into multiple sub-sequences, called *same-timestamp* (*ST*) *sub-sequences*, each of which encloses all the events (in the sensor event sequence) with the same timestamp; each ST sub-sequence constitutes, thus, the basic coding unit of the proposed solution.

Each ST sub-sequence is then ordered in increasing order of the largest spatial coordinate and represented by four data structures (DSs), i.e., two DSs containing the event spatial (location) information ($x$ and $y$ coordinates, respectively), one DS containing the polarity information and one DS containing the number of events that were triggered at the timestamp of the ST sub-sequence (which corresponds to the ST sub-sequence length). Each DS is subsequently encoded, where binarization is employed to encode the polarity information DS and the TTP algorithm is employed to predictively encode the other three DSs.

The TTP algorithm uses a short-depth decision tree based on a triple threshold to partitioning the range of the input data (notably $x$ and $y$ coordinates of the events and the ST sub-sequence length) into several smaller coding ranges distributed at concentric distances from the prediction, obtained from the corresponding previously coded data. Afterwards, each TTP input data value is represented by the binary representation of the prediction error and the binary representation of the decision tree structure, generating the corresponding elementary data structure encoded bitstream; the *event coding bitstream* is then formed by multiplexing the 4 elementary encoded bitstreams resulting from encoding the 4 DSs used to represent every ST sub-sequence.

In [32], it is also proposed a coding solution providing random access (RA) functionality, called *LLC-ARES-RA*. In LLC-ARES-RA, the sensor event sequence is first divided into multiple so-called *packages* of a given time length, which constitute the RA units. Each package is then coded with the LLC-ARES solution, and, to the resulting package encoded bitstreams, it is added a header information bitstream resulting from employing the TTP algorithm to code the length of package bitstreams.

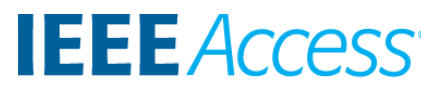

Experimental results on the DSEC dataset show that LLC-ARES has, on average, higher compression performance, measured in terms of CR, compared to the lossless benchmark solutions Bzip2 (lossless compression method based on Burrows–Wheeler algorithm), LZMA, and ZLib, with average CR improvements of 5.49%, 11.45%, and 35.57%, respectively; for event sequences with high density (i.e., with high number of events per second), Bzip2 and LZMA perform better than LLC-ARES, with an average CR up to 1.2× and 1.1× higher, respectively. As far as the LLC-ARES-RA solution is concerned, experimental results show a performance close to the LLC-ARES for the smallest (package) time length considered (100$\mu$s).

#### 11) FEATURE REPRESENTATION AND COMPRESSION METHODS FOR EVENT-BASED DATA [33]

In 2023, Wang et al. proposed a conceptually different lossless coding approach, which is based on a character-like representation of the event representation components (i.e., $\langle x, y, p, t\rangle$), to code event data generated from a DAVIS event camera (consisting of location, polarity, and timestamp) [33].

The proposed approach includes two coding methods, namely the *Characteristic Parameter Jointed Coding* (*CPJC*) method and the *ASCII Coding based on Bit Operation* (*ACBO*) method. Both methods are applied at the event level of the sensor event sequence and are based on the conversion of the $\langle x, y, p, t\rangle$ event representation into a sequence of characters (i.e., character-like representation); the event constitutes, thus, the basic coding unit of both proposed methods.

In the CPJC method, the event polarity, '0' or '1', is first mapped to the special characters '-' or 'null', respectively. Then, the pixel coordinates $x$ and $y$, i.e., the event location, are converted to an alphabetic character (letter), represented by 2 digits, followed by a numeric character, represented by 1 digit; the 3-digit representation of the $x$ and $y$ coordinates results from the fact that, in a DAVIS346 sensor with 346 × 260 spatial resolution, the maximum coordinate value, 346, requires 3 digits to be represented. In the proposed $x$ coordinate conversion process, the numeric character corresponds to the last (rightmost) digit of the $x$ coordinate value while the alphabetic character is obtained from a predefined dictionary, which associates a 2-digit value (the two leftmost digits of $x$ coordinate value) to a letter; a similar conversion process is applied to the $y$ coordinate value. Finally, the event timestamp $t$ is first subtracted from the timestamp of the previously triggered event and the difference is then represented by a numeric character. After applying the CPJC method to every event of the sensor event sequence, the corresponding sequence of characters are multiplexed and Zip compression is applied as entropy coding, generating the *event coding bitstream*.

In the ACBO method, the pixel coordinate $x$ (respectively, $y$) of every event in the sensor event sequence is first converted to a binary sequence, whose length is stored in a structure for posterior entropy coding, and the binary sequences associated to the $x$ (respectively, $y$) coordinate of all events in the sensor event sequence are then concatenated forming a long binary sequence. Next, the long binary sequence is split into multiple 7-bit-long binary sub-sequences, each of which is then converted to its corresponding decimal value for posterior ASCII character representation according to a predefined fine-tuned ASCII table; Zip compression is then applied to the corresponding sequence of ASCII characters as entropy coding, generating the event $x$ (respectively, $y$) coordinate encoded bitstream. The event timestamp $t$ is converted to a numeric character, as in the CPJC method, while the event polarity, '0' and '1', is represented by the '-' and '+' characters, respectively; the character representation of the polarity and timestamp information of each event are concatenated, with the polarity character acting as the separator between the timestamp characters of two consecutive events. Zip compression is then applied to the corresponding sequence of characters, generating the event polarity-timestamp encoded bitstream. The (full) *event coding bitstream* in the ACBO method results from the multiplexing the elementary (event) $x$ and $y$ coordinate encoded bitstreams with the polarity-timestamp encoded bitstream.

Experimental results on four event sequences, acquired by the authors with the DAVIS346 event camera, show that the proposed coding methods have higher compression performance, measured in terms of CR, compared to the lossless benchmark solution in [21] and direct Zip coding of the $\langle x, y, p, t\rangle$ event representation; CR improvements of 17.93% and 14.92% compared to [21] are reported for CPJC and ACBO methods, respectively. It is also observed that, on average, the CPJC method performs better than the ACBO method.

#### 12) MEMORY-EFFICIENT FIXED-LENGTH REPRESENTATION OF SYNCHRONOUS EVENT FRAMES FOR VERY-LOW-POWER CHIP INTEGRATION [34]

Also in 2023, Schiopu and Bilcu proposed a low complexity coding framework based on a memory-efficient fixed-length representation using multi-level lookup tables (LUTs) to code event frames created from event data generated from a DVS-like camera (consisting of location, polarity, and timestamp) while targeting to be suitable for hardware implementation in very low power event signal processing chips [34].

As in [28] and [30], the polarity of the events triggered at each pixel location are first summed up over a fixed time interval $\Delta$ and the polarity sum's sign $\{-1, 0, 1\}$ is then stored in the corresponding pixel location in the so-called *event frame* (*EF*), which constitutes the basic coding unit of the proposed framework. Hence, similarly to the solutions 6) and 8) reviewed above ([28] and [30], respectively), the coding framework proposed in [34] is classified as lossy/lossless, since, depending on the $\Delta$ value, it may involve event timestamp quantization and polarity accumulation, two processes that lead to some precision loss; these are, however, the only two operations inducing loss of information in the proposed coding framework.

Each EF is then partitioned in blocks of 32 × 32 pixels, each of which is vectorized and further split into 205 subsets of 5 ternary symbols $\{-1, 0, 1\}$. Each subset, i.e., 5 ternary symbols, is then remapped into an 8-bit symbol through a mathematical model and stored in a 205-symbol-long vector. A fixed-length LUT-based encoding solution is then applied to every 205-symbol-long vector, where a LUT-based representation is used to store all the unique combinations of 205 symbols found in any 205-symbol-long vector (obtained from the EF) and an index matrix is used to store in each entry the position, i.e., index, in the LUT-based representation from where the corresponding 205-symbol-long vector can be extracted. The index matrix and the LUT-based representation are then encoded through binarization, generating the *event coding bitstream*.

Experimental results on the DSEC dataset show that, for time intervals $\Delta = 1000\mu s$ and $\Delta = 5555\mu s$, the compression performance of the proposed solution, measured in terms of the aggregation CR (see Section V), approaches (but is still below) the performance obtained with the conventional lossless video/image coding benchmarks. For time intervals $\Delta = 1000\mu s$ ($10^3$fps) and $\Delta = 5555\mu s$ (180 fps), [34] reports average aggregation CR reductions up to 5.79%, 12.76%, 5.13%, and 33.27% compared to HEVC, VVC, CALIC, and FLIF, respectively; compared to the event-based coding benchmark in [30], an average aggregation CR reduction up to 46.85% is reported for the same time intervals. For the smaller time intervals, $\Delta = 1\mu s$ ($10^6$fps) and $\Delta = 100\mu s$ ($10^4$fps), it is possible to observe significantly higher average aggregation CR performance reductions when compared to the same benchmarks; [34] reports average aggregation CR reductions up to 81.92%, 79.42%, 86.57%, 88.83%, and 94.40% compared to HEVC, VVC, CALIC, FLIF, and [30], respectively. Please recall that, as in [28] and [30], the input to the conventional image/video coding solutions are (event) images obtained by combining, through some mathematical function, the information enclosed in each group of 5 (consecutive) EFs (please refer to [28] for more details).

#### 13) ENTROPY CODING-BASED LOSSLESS COMPRESSION OF ASYNCHRONOUS EVENT SEQUENCES [35]

Still in 2023, Schiopu and Bilcu extended the low complexity lossless coding framework in [32] (denominated LLC-ARES), by modifying the TTP algorithm to employ entropy coding based techniques, to code the *same-timestamp* (*ST*) *representation* of event data generated from a DVS-like sensor (consisting of location, polarity, and timestamp) [35].

In the proposed framework, named *Entropy coding-based Lossless Compression of ARES* (*ELC-ARES*), after ordering a ST sub-sequence in increasing order of the largest spatial coordinate, a triple threshold-based range partition (TTP) algorithm employing a set of adaptive Markov models (AMMs) is applied to predictively encode the four data structures (DSs) in which a ST sub-sequence is represented, i.e., the two DSs containing the event spatial (location) information ($x$ and $y$ coordinates, respectively), the DS containing the polarity information and the DS containing the number of events that were triggered at the timestamp of the ST sub-sequence; the Laplace estimator is used to compute the probability distribution for the TTP algorithm.

Different from the LLC-ARES solution [32], ELC-ARES adopts also a new $x$-coordinate prediction strategy, where the prediction of any element within the DS storing the $x$-coordinate data corresponds to the $x$-coordinate associated to the first event triggered at the ST sub-sequence timestamp; as a consequence, ELC-ARES also adopts a new initialization for the triple threshold needed for the TTP algorithm associated to the $x$-coordinate. In addition to the 4 DSs coding processes, a TTP algorithm similar to the one described above (i.e., employing a set of AMMs) is also applied to encode the decision trees resulting from the coding processes of all DSs except the polarity related one; the *event coding bitstream* is then formed by multiplexing the elementary bitstreams resulting from encoding the 4 DSs and the decision trees.

Experimental results on the DSEC dataset show that ELC-ARES has higher compression performance, measured in terms of CR, compared to the lossless benchmark solutions LLC-ARES, Bzip2, LZMA and ZLib, with average CR improvements (over the DSEC dataset) of 21.40%, 28.03%, 35.27%, and 64.54% respectively. Experimental results also show that ELC-ARES provides *average event encoding speed* reductions of 21.41%, 91.10%, and 54.74% compared to Bzip2, LZMA and ZLib, respectively, and the LLC-ARES benchmark solution achieves an *average event encoding speed* 46% lower than the proposed ELC-ARES solution; the *average event encoding speed* metric measures the (average) time needed to encode one event (see TABLE 4).

#### 14) EVENT DATA STREAM COMPRESSION BASED ON POINT CLOUD REPRESENTATION [36]

In 2023, Huang and Ebrahimi proposed a lossless coding solution based on a point cloud representation to code event data generated from a DAVIS event camera (consisting of location, polarity, and timestamp) [36].

Similarly to [31], the sensor event sequence is organized as a set of points in 3D (space-time) volume, where the pixel location $(x, y)$ and the timestamp $t$ define the 3D coordinate axes $X$, $Y$ and $Z$, respectively, and the polarity is the value attributed to each 3D point, thus resembling a 3D point cloud representation. In this context, the events of the sensor event sequence are first aggregated, according to their polarity, into multiple sets with a fixed number of events each, generating multiple sets of two 3D point clouds (one for each polarity), all with the same number of events. Next, the $(x, y, t)$ values of each point in each point cloud are scaled to a range of values in the $x$, $y$, and $t$ coordinates appropriate for coding; in this case, timestamp $t$ is multiplied by a temporal scaling factor of $1 \times 10^6$ and $x$ and $y$ coordinates are multiplied by a special scaling factor of $1 \times 10^3$. Then, the standard Geometry-based Point Cloud Compression (G-PCC) codec [41] is applied to encode the geometry information (i.e., the $(x, y, t)$ triplet) of every

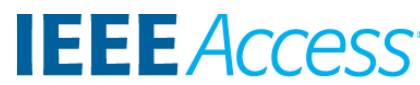

3D (polarity-based) point cloud separately. The elementary encoded bitstreams resulting from applying G-PCC to every 3D (polarity-based) point cloud are then multiplexed to generate the *event coding bitstream*.

Experimental results on 8 (indoor and outdoor) event sequences of the DAVIS 240C dataset (only event data considered) show higher compression performance, measured in terms of CR, compared to the lossless benchmark solutions [21], LZMA, Sprintz-Delta (compression algorithm for Internet of Things devices characterized by low memory consumption and low latency), Huffman coding, and SIMD-BP128 (vectorized binary packing coding scheme), with average CR improvements (over the 8 event sequences) of 22.2%, 36.2%, 100.4%, 161.4%, and 264.9% respectively.

#### 15) A NOVEL APPROACH FOR NEUROMORPHIC VISION DATA COMPRESSION BASED ON DEEP BELIEF NETWORK [37]

Also in 2023, Khaidem et al. proposed a deep learning-based coding framework to code pseudo video sequences created from event data generated from a DAVIS event camera (consisting of location, polarity, and timestamp) [37].

Similarly to [25], the proposed solution first accumulates, over a fixed time interval, the number of events triggered at each pixel location according to their polarity, generating (two) polarity-based *event frames*; these frames have the full sensor pixel array spatial resolution and represent, at each pixel location, the event count for a given polarity, i.e., correspond to location histograms. As for solution 4) reviewed above ([25]), the coding framework proposed in [37] is classified as lossy since it requires event timestamp quantization, an operation that induces some precision loss in the timestamp component.

The two polarity-based event frames created from event aggregation are then concatenated side by side creating a so-called *superframe,* i.e., a frame with twice the sensor pixel array horizontal resolution. This process is repeated over the whole sensor event sequence duration and the resulting set of superframes, structured into a 3D array of superframes, is then treated as a pseudo video sequence, with each superframe constituting the basic coding unit of the proposed framework.

The proposed superframe coding process starts by dividing each superframe into 30 × 30 blocks and feeding them to a deep belief network (DBN), comprising a 4-layer autoencoder, generating low-dimensional (20 × 1) latent features vectors. The resulting latent features vectors are then encoded using Huffman arithmetic coding, generating the corresponding (superframe blocks) coding bitstreams. The elementary bitstreams resulting from encoding all 30 × 30 blocks of all superframes are finally multiplexed, forming the *event coding bitstream*.

Experimental results on 3 (indoor) sequences of the DAVIS 240C dataset (only event data considered) show significant compression performance gains in general, measured in terms of CR versus aggregation time interval, with respect to several benchmark solutions, such as the event coding solutions in [21] (lossless solution) and [25] (lossy solution), and the generic lossless data compression algorithms Huffman arithmetic coding, LZMA, LZ4, ZLib (Zeta Library), Zstd, Brotli, and Snappy (fast integer compression algorithm). For the largest time aggregation interval considered (30ms), the proposed solution achieves an average CR (over the 3 event sequences) 44.35× higher than the one achieved with the lossless event coding solution in [21] and an average CR up to 108.95× higher than the ones achieved with the generic data compression algorithms; the average CR improvements decrease as the time aggregation interval decreases. However, it is worth mentioning that, according to [37], the DBN was trained on blocks derived from the first 10 seconds of the event sequences used for validation, which may bias the results and somehow justify the high performance obtained.

#### 16) LOSSLESS ENCODING OF TIME-AGGREGATED NEUROMORPHIC VISION SENSOR DATA BASED ON POINT-CLOUD COMPRESSION [38]

In 2024, Adhuran et al. extended the TALVEN solution in [25], by proposing to use a standard point cloud compression scheme to code the time-aggregated event data created from event data generated from a DAVIS event camera (consisting of location, polarity, and timestamp) [38].

Similarly to [25], the proposed solution, so-called *Time-Aggregated Lossless Encoding of Events based on Point-Cloud Compression* (TALEN-PCC), first accumulates, over a fixed time interval $\Delta$, the number of events triggered at each pixel location according to their polarity. This process generates (two) polarity-based event matrices, each with the full spatial resolution of the sensor's pixel array. These matrices correspond to location histograms, where each pixel location reflects the event count for a given polarity. As for solution 4) reviewed above ([25]), the coding framework proposed in [38] is classified as lossy since it requires event timestamp quantization, an operation that induces some precision loss in the timestamp component.

The two polarity-based event matrices created from event aggregation are then (independently) raster scanned and the matrices entries with non-zero event counts are arranged in two so-called *multivariate streams* (one for each polarity), where each non-zero matrix entry is represented by 4 variables: $x$, $y$, *Event Count* and $k\Delta$; $(x, y)$ corresponds to the pixel location and $k\Delta$ corresponds to the aggregation time interval index in which the *Event Count* events occurred. This event accumulation and multivariate streams creation process is repeated over the whole duration of the sensor event sequence (i.e., raw input event sequence). Each resulting multivariate stream constitutes, thus, the basic coding unit of the proposed solution.

Each resulting multivariate stream is organized as a 3D point cloud, i.e., a set of points in 3D (space-time) volume, where the pixel location $(x, y)$ and the aggregation time interval index $k\Delta$ correspond to the 3D coordinate axes $X$, $Y$ and $Z$, respectively, and the Event Count is the attribute associated with each 3D point. Each 3D point cloud is then encoded with the standard Geometry-based Point Cloud Compression (G-PCC) codec [41]; while $x$, $y$, and $k\Delta$ are encoded as geometry information, the Event Count is encoded as attribute information (i.e., reflectance input of the G-PCC codec). The elementary encoded bitstreams (geometry and attribute) resulting from applying G-PCC to every 3D point cloud (i.e., multivariate stream) are then multiplexed to generate the *event coding bitstream*.

Experimental results on the DAVIS 240C dataset (only event data considered), show higher coding performance, measured in terms of CR, when compared to the TALVEN solution in [25], achieving CRs up to 30% higher. Additionally, for medium to high aggregation time intervals, the TALEN-PCC method significantly outperforms the lossless event coding solution in [21].

#### 17) FLOW-BASED VISUAL STREAM COMPRESSION FOR EVENT CAMERAS [39]

Also in 2024, Stumpp et al. proposed the so-called *flow-based compression* (FBC), a conceptually different lossy coding approach, called *stream-to-stream* approach. Unlike the traditional coding approach adopted by all the remaining reviewed NVDC solutions, where the encoder output is a coding bitstream, the proposed FBC solution encodes the sensor event sequence into another (reconstructed) event sequence. The FBC encoding operation is based on not transmitting future events for some time period; instead, those future events are predicted at the receiver using real-time optical flow estimates computed and sent by the transmitter. The proposed FBC approach is designed specifically for real-time compression of event data generated from a DAVIS event camera (consisting of location, polarity, and timestamp) [39].

The proposed FBC solution operates in two phases: a *sending phase* followed by a *predicting phase*. During the time period of the sending phase, event-wise optical flow is first computed using the hARMS method [44]. Then, the events are classified into *flow events* or *no-flow events* based on the computed flow consistency. A *flow event* corresponds to an event with consistent flow that can be reliably used to perform prediction of future events at the receiver side. Flow events are transmitted along with their computed optical flow estimates. These flow estimates are used by the receiver to predict future events, allowing the FBC to reduce the actual number of events transmitted, thus leading to compression. A *no-flow event* corresponds to an event with a non-consistent flow. No-flow events are transmitted as they are, i.e., with 64 bits per event, without any additional compression.

During the time period of the predicting phase, the receiver uses each flow event received during the previous sending phase to predict the spatio-temporal location of future events; those future events are not transmitted, as they are assumed to be accurately predicted based on the consistency of the optical flow computed at the transmitter. This prediction process reduces the need to transmit every event, leading to compression. Once the predicting phase is concluded for every flow event received, the receiver combines the predicted events with the no-flow events (received during the predicting phase) generating the (FBC) reconstructed event sequence. While the proposed FBC approach targets to be suitable for real-time compression of event data, it can be combined with other compression strategies for improved performance [39], i.e., the (FBC) reconstructed event sequence can be further compressed.

Experimental results on 4 event sequences (obtained from the DAVIS 240C dataset, [45] and [46]) show an average CR of 2.8, corresponding to an average event reduction of 68%, with a median temporal error of 0.48ms and an average spatio-temporal event stream distance of 3.07, in a coding framework where only the proposed FBC is used; please refer to TABLE 4 for more details on the event reduction, temporal error and spatio-temporal event stream distance. By applying LZMA on top of the event sequence reconstructed by the proposed FBC, experimental results show a CR improvement of 3.72× with respect to the scenario where LZMA is not used.

### B. SPIKE-BASED NVDC SCHEMES

#### 1) AN EFFICIENT CODING METHOD FOR SPIKE CAMERA USING INTER-SPIKE INTERVALS [23]

In 2019, Dong et al. proposed the first lossy coding framework to code time interval sequences created from spike data generated from a spike camera (consisting of 'ON'/'OFF', i.e., binary, values) [23].

In the proposed framework, each *spike train*, i.e., each sequence of spikes ('ON'/'OFF' or '1'/'0' values) outputted by a single sensor pixel along time, is first converted into a sequence of 'waiting' times between consecutive spikes, i.e., relative latency of spikes, known as *inter-spike intervals* (*ISIs*). Then, each ISI sequence is adaptively partitioned into multiple sub-sequences (temporally), called *segments*, such that adjacent segments are characterized by different ISIs distributions and, thus, a more efficient exploitation of the spatio-temporal redundancies can be achieved; each (ISI) segment (from a sensor pixel) constitutes, thus, the basic coding unit of the proposed framework, with ISI corresponding to the time interval component of the spike representation targeted by coding (see Section II-B).

The proposed (ISI) segment encoding process involves evaluating two intra-pixel prediction modes, the so-called *mean value mode* (*MVM*) and *forward mode* (*FM*), and one inter-pixel prediction mode, and selecting the mode leading to the lowest rate-distortion cost; considering that the average pixel intensity is inversely proportional to the ISI, an intensity-based distance between two spike trains is proposed as distortion measure. While *intra-pixel prediction* considers ISI data only from segments belonging to the pixel to be encoded, *inter-pixel prediction* considers ISI data from segments belonging to both the pixel to be encoded and to neighboring pixels; the spike location component is somehow embedded in the order in which the basic coding units (each segment from a sensor pixel) are scanned and, thus, it is not directly coded.

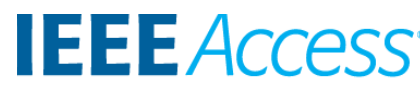

The intra-pixel prediction mode MVM, designed to deal with homogeneous segments, i.e., segments with similar ISI values, involves the computation of the mean value of the ISIs within a segment followed by the subtraction of that mean value from all the segment ISI values, obtaining ISI (prediction) residuals. The intra-pixel prediction mode FM, designed to deal with non-homogeneous segments, i.e., segments with varying ISI values, involves motion estimation and compensation in the temporal dimension only, i.e., considering previously coded segments from the same pixel location only. After finding the reference (segment) candidate that minimizes the intensity-based distance with respect to the segment to be encoded, the (prediction) residuals of ISIs are obtained and the associated motion vector is predicted from the average of the motion vectors of previously coded segments with the FM mode.

As far as the inter-pixel prediction mode is concerned, it involves motion estimation and compensation in both temporal and spatial dimensions, i.e., considering previously coded segments from both the same pixel and neighboring pixels in a causal spatio-temporal window. Thus, for each reference (segment) candidate, the intensity-based distance with respect to the segment to be encoded is computed and the one with the lowest distance is selected as the prediction segment; the (prediction) residuals of ISIs are then obtained by subtracting the prediction segment from the segment to be encoded. The spatio-temporal motion vector associated to the prediction segment (with components in $x$, $y$, and $t$) is then predicted from the average of the spatio-temporal motion vectors of previously coded segments from the same pixel and from segments of 4 neighboring pixels (in a causal spatial window). After the intra- and inter-pixel coding, the corresponding prediction residuals are quantized (with a varying quantization step size at each ISI) and context-based adaptive entropy encoded, generating the *spike coding bitstream*.

Experimental results on the PKU-Spike dataset, proposed in [23] for the spike data coding algorithm evaluation, show compression performance, measured in terms of CR for several quantization parameter (QP) values, with average CR values (over the PKU-Spike dataset) ranging between 23.04 (for QP = 4) and 53.41 (for QP = 32). It is also shown in [23] the PSNR and SSIM evolution with QP, where each PSNR/SSIM value corresponds to the average of 1000 PSNR/SSIM values computed for 1000 still intensity images reconstructed from raw spike data and decoded spike data.

#### 2) HYBRID CODING OF SPATIOTEMPORAL SPIKE DATA FOR A BIO-INSPIRED CAMERA [26]

In 2021, Zhu et al. extended the lossy coding framework in [23], by incorporating an adaptive polyhedron partitioning, intra and inter polyhedron-based prediction, transform and multi-layer quantization, to code time interval sequences created from spike data generated from a spike camera (consisting of 'ON'/'OFF', i.e., binary, values) [26].

In the proposed framework, each spike train (sequence of 'ON'/'OFF' spikes) outputted by a single sensor pixel along time, is first converted into a sequence of inter-spike intervals (ISIs), i.e., 'waiting' times between subsequent spikes. Then, the ISI sequences associated to all the sensor pixels are structured in an ISI volume and divided into *macro cuboids*, i.e., cuboids with the full sensor pixel array spatial resolution and a predefined time length. Each macro cuboid is in turn partitioned into multiple *spike cuboids*, each of which corresponding to $2 \times 2$ pixels in the spatial dimension, called *pixel group*. The set of ISI sequences belonging to a pixel group are further adaptively partitioned into multiple polyhedrons according to the motion characteristics; the polyhedron constitutes, thus, the basic coding unit of the proposed coding framework.

Intra and Inter polyhedron-based prediction with *spike-rate* and *spike-time* modes (4 prediction modes in total) are then evaluated and the prediction mode leading to the lowest rate-distortion cost is selected as the best prediction mode; an intensity-based distance between two spike trains is proposed as distortion measure. While the Intra polyhedron-based prediction modes (Intra spike-rate and Intra spike-time) consider only the spike data of the ($2 \times 2$) pixel group in the polyhedron, the Inter polyhedron-based prediction modes (Inter spike-rate and Inter spike-time) perform motion estimation in both temporal and spatial dimensions, i.e., considering previously coded polyhedrons of neighboring spike cuboids that are nearest to the polyhedron to be coded.

The *Intra spike-rate mode*, designed to deal with simple (regular) motion patterns or static regions, involves estimating the spike firing rate of each pixel in the ($2 \times 2$) pixel group in the polyhedron and use the central pixel firing rate to predict the firing rates of the remaining pixels in the pixel group. The *Intra spike-time mode*, designed to deal with motion patterns slightly more complex than the ones targeted by the Intra spike-rate mode, in addition to encode the firing rate of each pixel in the polyhedron to be coded (as in the previous Intra mode), it involves using the estimated spike firing rate of each pixel in the ($2 \times 2$) pixel group to reconstruct the spike data and obtaining the prediction residuals of ISIs by subtracting the reconstructed spike data from the one to be encoded.

The *Inter spike-rate mode*, designed also to deal with static regions, extends the Intra spike-rate prediction mode by performing motion estimation and compensation in both temporal and spatial dimensions, i.e., considering previously coded polyhedrons from both the current spike cuboid and from the 4 neighboring spike cuboids in a causal spatio-temporal window. Thus, for each reference (polyhedron) candidate, the intensity-based distance with respect to the polyhedron to be encoded is computed and the one with the lowest distance is selected as the reference polyhedron. Next, the polyhedron to be encoded is predicted from the spike firing rate of each pixel in the reference polyhedron, and the prediction residual of the spike firing rate is encoded as described in the Intra spike-rate mode. The *Inter spike-time mode*, designed to deal with complex motion patterns, involves also motion estimation and compensation in both temporal and spatial dimensions as in the Inter spike-rate mode, but considers a wider spatio-temporal (cuboid) window where motion search is performed. Thus, for each polyhedron to encode, the best match to the spike data (ISIs) associated to each pixel in the (2 × 2) pixel group in polyhedron is searched for in the spatio-temporal (cuboid) window, by minimizing the sum of the intensity-based distances between every pixel spike data (ISIs) and the matching spike data (ISIs), and the corresponding (2 × 2) spatio-temporal motion vectors are obtained. The prediction residual of ISIs is then obtained by subtracting from the ISIs of each pixel in the (2 × 2) pixel group in the polyhedron to encode the corresponding data from the best match found. The spatio-temporal motion vectors are predicted from the average of the spatio-temporal motion vectors of previously Inter spike-time mode coded polyhedrons from the current spike cuboid and from the 4 neighboring spike cuboids in a causal spatio-temporal window.

While the prediction residuals of the spatio-temporal motion vectors are lossless encoded, the prediction residuals of the spike firing rate and the prediction residuals of the ISIs are (2D) DCT transformed. The DCT transformed coefficients of the spike firing rate and ISI are then quantized (with a varying quantization step size at each ISI) and context-based adaptive entropy encoded, generating the *spike coding bitstream*.

Experimental results on an extended version of the PKU-Spike dataset presented in [23] show that the polyhedron-based prediction provides better compression performance, measured in terms of CR versus spike train intensity-based distance, than the pixel segment-based prediction proposed in [23]. Experimental results also include compression performance, measured in terms of CR, for several quantization parameter (QP) values, with average CR values (over the PKU-Spike dataset) ranging between 91.46 with an average distortion of 8.56 (for QP = 4) and 840.06 with an average distortion of 33.88 (for QP = 60); distortion corresponds to an intensity-based distance between two spike trains (see TABLE 4). The compression performance of the proposed coding framework is also compared with H.264/AVC video coding (main profile) for 1000 still intensity images reconstructed from the raw spike data and decoded spike data, considering CR as a function of PSNR and as a function of intensity-based distance between two spike trains (proposed in [26]). From the small set of results reported, H.264/AVC seems to be the better performing coding solution (by a large margin) for high-speed sequences and for lower CRs of static sequences, as far as the CR versus PSNR comparison is concerned. When it comes to the CR versus intensity-based distance comparison, the proposed coding framework always outperforms H.264/AVC, due to some information (e.g., firing time of spikes) loss occurring when converting spike data into intensity images.

## IV. NVDC: DATASETS OVERVIEW

In sections III and II-C, the NVDC solutions currently available in the literature have been reviewed and classified at the light of the proposed taxonomy, respectively, to exercise and demonstrate its potential. It is well known that, for a fair and straightforward comparison of coding solutions performances, it is essential to select meaningful and precise test conditions, among which the test material (or datasets) is a key part; the definition of meaningful and precise test conditions (and performance evaluation metrics) is, in fact, one of the short term goals of JPEG XE, a recent JPEG exploration activity on event-based vision [19].

In this context, this section provides an overview, in the form of a summary table (see TABLE 2), of the datasets used to evaluate the performance of the currently available NVDC solutions (reviewed in Section III). Besides associating to each reference the respective dataset(s) used, TABLE 2 also includes the datasets URLs, a brief description of the datasets content and some specs on the camera/sensor used to acquire the dataset (model, resolution and setup). The Raw Data Type classification dimension (of the proposed taxonomy) is also included in TABLE 2 for a faster identification of the datasets used by NVDC solutions sharing the same type of raw data at its input. As in Section III, the datasets presentation in TABLE 2 is grouped by the Raw Data Type classification dimension and follows the chronological order of the references in which they appear within each Raw Data Type class. As for TABLE 1, '?' indicates that not enough information could be found in the reference to clarify the respective entry. Moreover, the acronyms 'ATIS', 'LCD', 'RTK', 'GPS', 'IMU' in TABLE 2 stand for *Asynchronous Time-based Image Sensor*, *Liquid Crystal Display*, *Real-Time Kinematic*, *Global Positioning System* and *Inertial Measurement Unit*, respectively, and 'res.' abbreviation stands for *resolution*. Please recall that this section does not intent to list all the publicly available neuromorphic vision datasets but rather to identify the ones that have been used so far in the performance assessment of the NVDC solutions available in the literature (and reviewed in Section III).

TABLE 3 provides a concise and complementary view of datasets listed in TABLE 2, summarizing the datasets main characteristics while adding other relevant information of the neuromorphic vision data to the coding scenario, such as the event/spike sequences duration, number of events/spikes per sequence, etc. As for the previous tables, the symbol '?' in TABLE 3 indicates information not available in the references or that is not enough to clarify the respective entry while the symbol '-' indicates that the entry is not applicable to the respective dataset. In TABLE 3, 'APS', 'ADAS', SLAM, 'cam', 'avg.', 'train.' and 'val.' stand for *Active Pixel Sensor*, *Automatic Driver Assistance Systems*, *Simultaneous Localization And Mapping*, *camera*, *average*, *training*, and

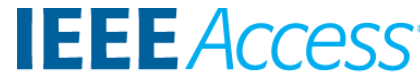

**TABLE 2.** Overview of the datasets used in currently available NVDC literature. Datasets URLs included in the table were accessed on May 2, 2024.

| Refs. using Dataset | Raw Data Type | Dataset | Dataset URL | Brief Contents Description | Camera/Sensor (*spatial res.*) | Camera/ Sensor Setup |
|---|---|---|---|---|---|---|
| [21] | Event | PKU-DVS (*PeKing University - Dynamic Vision Sensor dataset*) | https://pkuml.org/resources/pku-dvs.html | - Event imaging system, consisting of a static event camera, collecting data from real moving objects, persons, and vehicles;<br>- Data captured under different conditions, including daytime, night, close view, distant view, low-speed, and high-speed motion scenarios;<br>- 13 event sequences in total;<br>- 2 classes: indoor (7 seqs.) and outdoor (6 seqs.);<br>- Events captured with 240×180 ? spatial resolution. | DAVIS240B ?<br>(*240×180 ?*) | Static |
| [22] | Event | DDD17 (*DAVIS Driving Dataset 2017*) | https://sensors.ini.ch/datasets#h.6hvtcavazphj | - Event imaging system, consisting of an event camera mounted behind the windshield, collecting data from real highway and city driving scenarios;<br>- Data captured under several weather, driving, road, and lighting conditions;<br>- 40 event+image sequences in total;<br>- Events and intensity images (@ 24 Hz) captured with 346×260 spatial resolution;<br>- Dataset also contains IMU measurements and vehicle control-related data such as vehicle speed, steering angle, GPS position, etc. | DAVIS346B<br>(*346×260*) | Static & Moving |
| [24], [29] | Event | MNIST-DVS (*Modified National Institute of Standards and Technology - Dynamic Vision Sensor dataset*) | http://www2.imse-cnm.csic.es/caviar/MNISTDVS.html | - Event imaging system, consisting of an event camera placed in front of a 19" LCD monitor, collecting data from slowly moving handwritten digit images displayed on the LCD monitor;<br>- 30000 event sequences in total, obtained from 10000 original 28×28 pixel handwritten digit images from the MNIST dataset† scaled to three different sizes (4, 8, and 16);<br>- 10 classes: digit 0 to 9 (3000 seqs. per class);<br>- Events captured with 128×128 spatial resolution. | DVS128<br>(*128×128*) | Static |
| [25], [27], [31], [36], [37], [38], [39] | Event | DAVIS 240C (*Event-camera dataset*) | https://rpg.ifi.uzh.ch/davis_data.html | - Event imaging system, consisting of a moving event camera, collecting data from simple shapes to complex structures;<br>- 25 event+image sequences in total;<br>- 2 classes: indoor (22 seqs.) and outdoor (3 seqs.);<br>- Events and intensity images (@ 24 Hz) captured with 240×180 spatial resolution;<br>- Dataset also contains IMU measurements and camera calibration from the DAVIS camera as well as ground truth measurements (pose and orientation) from a motion-capture system. | DAVIS240C<br>(*240×180*) | Moving |
| [28], [30], [32], [34], [35] | Event | DSEC (*Stereo Event Camera dataset for Driving scenarios*) | https://dsec.ifi.uzh.ch/ | - Stereo (frame-based) RGB and event camera imaging system, consisting of two monochrome event cameras and two global shutter color cameras mounted on top of a car, collecting data from real driving in a variety of illumination conditions, including daytime and night;<br>- 53 stereo event-image sequences in total;<br>- 2 test sets: training (41 stereo seqs.) and test (12 stereo seqs.);<br>- Events captured with 640×480 spatial resolution;<br>- Color intensity images captured @ 20 Hz with 1440×1080 spatial resolution;<br>- Dataset also contains ground truth depth maps (for event-based stereo matching algorithms), Lidar data and RTK GPS measurements (for accurate positioning). | 2× Prophesee Gen3.1<br>+<br>2× color FLIR Blackfly S USB3<br>(*Event: 640×480,*<br>*RGB: 1440×1080 @ 20 Hz*) | Moving |
| [29] | Event | DvsGesture (*Dynamic vision sensor based hand and arm gestures dataset*) | https://research.ibm.com/publications/a-low-power-fully-event-based-gesture-recognition-system | - Event imaging system, consisting of a static event camera, collecting data from 29 subjects performing hand and arm gestures against a static background under 3 illumination conditions (combinations of natural light, fluorescent light, and LED light);<br>- 1342 event sequences in total, with a duration of ~6s each (gesture) sequence: 122 (gesture) sequences per class;<br>- 2 test sets: training (1078 seqs. from 26 subjects) and validation (264 seqs. from 3 subjects);<br>- 11 classes: 11 hand and arm gesture different types;<br>- Events captured with 128×128 spatial resolution;<br>- Dataset also contains frame-based (RGB) videos from a webcam mounted close to the event camera, and ground truth data with gesture labels and start and stop times. | DVS128<br>(*128×128*) | Static |

**TABLE 2.** ***(Continued.)*** **Overview of the datasets used in currently available NVDC literature. Datasets URLs included in the table were accessed on May 2, 2024.**

| | | | | | | |
|---|---|---|---|---|---|---|
| [29] | Event | Gen1 N-CARS<br>(*Large real-world event-based car classification dataset*) | http://www.prophesee.ai/dataset-n-cars/ | - Event imaging system, consisting of an event camera mounted behind the windshield of a car, collecting data from several real driving sessions;<br>- 24029 event sequences in total, with a 100ms duration each sequence;<br>- 2 classes: car (12336 seqs.) and background (11693 seqs.);<br>- 2 test sets: training (7940 car seqs. and 7482 background seqs.) and test (4396 car seqs. and 4211 background seqs.);<br>- Events are captured with 304×240 spatial resolution. | Prophesee Gen1<br>(aka ATIS)<br><br>(*304×240*) | Moving |
| [29] | Event | Gen1 Automotive Detection<br>(*Large scale event-based detection dataset for automotive*) | https://www.prophesee.ai/prophesee-gen1-automotive-detection-dataset/ | - Event imaging system, consisting of an event camera mounted on a car dashboard, collecting data from several real driving scenarios;<br>- Data captured considers open roads and diverse driving scenarios, including urban, highway, suburbs, and countryside scenes, under different weather and illumination conditions;<br>- 2359 event sequences in total, with a 60s duration each sequence;<br>- 3 test sets: training (1460 seqs.), test (470 seqs.) and validation (429 seqs.);<br>- Events are captured with 304×240 spatial resolution;<br>- Dataset also contains 255781 manual bounding box annotations of cars (228123) and pedestrians (27658). | Prophesee Gen1<br>(aka ATIS)<br><br>(*304×240*) | Moving |
| [33] | Event | *not available* | *not available* | - Event imaging system, consisting of a static event camera, collecting data from real shapes and scenes under different motion characteristics;<br>- 4 event sequences in total;<br>- Events captured with 346×260 spatial resolution. | DAVIS346<br><br>(*346×260*) | Static |
| [23], [26] | Spike | PKU-Spike-A<br>(*PeKing University - Spike camera dataset*) | https://www.pkuml.org/resources/pku-spike.html | - Spike imaging system, consisting of a spike camera with time sampling frequency of 40000 Hz, collecting data from real objects/scenes;<br>- 6 spike sequences in total, with 3.64s duration each sequence;<br>- 2 classes: normal-speed motion (3 seqs.) and high-speed motion (3 seqs.);<br>- Spikes captured with 400×250 spatial resolution. | Vidar<br><br>(*400×250*) | Static |

†http://yann.lecun.com/exdb/mnist/

*validation*, respectively. In 'Sequence Duration' and 'Total # Event/Spike per Sequence' columns, '∼' indicates approximately, while 'a – b' indicates *minimum – maximum* range of values; in 'Total # Event/Spike per Sequence' column, 'K', 'M' and 'G' stand for *kilo*, *mega,* and *giga*, respectively.

From TABLE 2 and TABLE 3, the following conclusions can be obtained:

- For the event Raw Data Type class, there are 2 event-based datasets (from a total of 9) that have been more often adopted in the event-based NVDC solutions performance assessment (including the most recent ones): the DAVIS 240 and the DSEC datasets; the DAVIS 240 dataset popularity seems, however, to be higher among the research groups around the world working on this emerging area (4 different research groups adopted the DAVIS 240 dataset while only 1 adopted the DSEC dataset). Besides being characterized by considerably different spatial resolutions (240 × 180 for both event data and monochrome intensity images versus 640 × 480 for event data and 1440 × 1080 color intensity images), these datasets also target different tasks, characterized by significantly different scene dynamics that translate into a different number of events per sequence; the number of events per sequence in the DSEC dataset is 16 times to 105 times higher than the one in the DAVIS 240 dataset. For the spike Raw Data Type class, a single spike-based dataset (with 400 × 250 spatial resolution) has been adopted in the performance assessment of both spike-based NVDC solutions.
- All neuromorphic vision datasets result from the acquisition (with a static and/or moving camera setup) of real-world scenes, although through a variety of different imaging systems since they address rather different tasks (from pose estimation and objects/actions recognition, detection, tracking and classification to simultaneous localization and mapping).

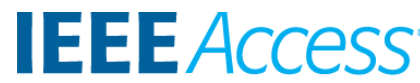

**TABLE 3. Summary of the main neuromorphic vision data related characteristics of the datasets used in currently available NVDC literature. Hyperlinks to datasets included in the table were accessed on May 2, 2024.**

| **Dataset (*Refs. using dataset*)** | **Task** | **Camera/Sensor Model** | **Spatial Resolution [pixels]** | **Acquisition** | **Total # Sequences** | **Sequence Duration [s]** | **Total # Event/Spike per Sequence [events/spikes]** | **# Classes (*# seqs./class*)** | **# Test Sets (*# seqs./test set*)** |
|---|---|---|---|---|---|---|---|---|---|
| PKU-DVS ([21]) | DVS-based coding solutions assessment | DAVIS240B ? [43] | 240×180 ? | Static cam., Real-world, Event data | 13 | 2.10 – 355.21 | ~113.7K – ~30.5M | 2 (*7 indoor, 6 outdoor*) | - |
| DDD17 ([22]) | Study of fusion of APS images and DVS data for ADAS | DAVIS346 [47][48] | 346×260 | Static & moving cam., Real-world, Event+grayscale image data, Vehicle control and GPS ground truth data | 40 | 10 – 3135 | 267 – ~340.1K | - | - |
| MNIST-DVS ([24], [29]) | Digits classification | DVS128 [49] | 128×128 | Static cam., Moving image on LCD, Event data | 30000 | ~2.5 | Scale 4: 17011 (avg.) Scale 8: 43764 (avg.) Scale 16: 103144 (avg.) | 10 (*3000/class*) | - |
| DAVIS 240C ([25], [27], [31], [36], [37], [38], [39]) | Pose estimation, visual odometry, and SLAM | DAVIS240C [50] | 240×180 | Moving cam., Real-world, Event+grayscale image data, Camera pose & orientation ground truth data | 25 | 3.4 – 133.4 | ~1.1M – ~185.7M | 2 (*22 indoor, 3 outdoor*) | - |
| DSEC ([28], [30], [32], [34], [35]) | Event-based stereo algorithms assessment | Gen3.1 [51] (*RGB cam.: FLIR Blackfly S USB3*) | 640×480 (*RGB cam: 1440×1080 @ 20 Hz*) | Moving cam., Real-world, Event+color image data, Depth & RTK GPS ground truth data | 53 stereo | ~11 – ~100 ? | ~115M – ~3G | - | 2 (*41 train., 12 test*) |
| DvsGesture ([29]) | Gesture recognition | DVS128 [49] | 128×128 | Static cam., Real-world, Event data | 1342 | ~6 (avg.) | ? | 11 (*122/class*) | 2 (*1078 train., 264 test*) |
| Gen1 N-CARS ([29]) | Car classification | ATIS [52] (aka Gen1.0) | Variable size crops from 304×240 | Moving cam., Real-world, Event data | 24029 | 0.1 | ~501.2 – ~50.1K | 2 (*7940 car, 7482 background / 4396 car, 4211 background*) | 2 (*15422 train., 8607 test*) |
| Gen1 Automotive Detection ([29]) | Automotive detection | ATIS [52] (aka Gen1.0) | 304×240 | Moving cam., Real-world, Event data | 2359 | 60 | ? | - | 3 (*1460 train., 470 test, 429 val.*) |
| *not available* ([33]) | *not available* | DAVIS346 [47][48] | 346×260 | Static cam., Real-world, Event+grayscale image data | 4 | ? | 3072000, 6886500, 7380625, 11717625 | - | - |
| PKU-Spike-A ([23], [26]) | Assess spike-based coding solutions | Vidar [5] | 400×250 | Static cam., Real-world, Spike data | 6 | 3.84 | ~125.6M – ~930.6M | 2 (*3 normal-speed, 3 high-speed*) | - |

- The ranges of the values of the number of events/spikes per sequence tend to have a high amplitude. This may be an indicator that the datasets cover some scene dynamics variety (as the scene dynamics directly impacts on the number of events/spikes generated by the neuromorphic vision camera/sensor), which is important to obtain representative and meaningful performance results.
- Four out of ten neuromorphic vision datasets also contain grayscale or color (intensity) images captured along with the neuromorphic vision data (in this case event data) acquisition. However, this conventional type of visual data (intensity images) has only been exploited in the event data coding process by one NVDC solution, the PDS-LEC solution [27].
- A few datasets, e.g., Gen1 NCARS and Gen1 Automotive Detection, appear to have already been designed with the prospective development of deep learning-based NVDC solutions in mind; those datasets have already event sequences for training and validation purposes.

In summary, the DAVIS 240 dataset appears to be the most popular dataset among the research community when it comes to the NVDC performance assessment. However, its content (see TABLE 2) hardly reflects the content of the NVDC main applications scenarios [20]. Thus, the definition and adoption of reference neuromorphic vision dataset(s) reflecting the content variety of the NVDC main applications scenarios (together with precise test conditions, performance evaluation metrics, and benchmarking coding solutions) is, therefore, of utmost importance to promote solid and consistent advancements in this emerging technical area. Given the recent JPEG exploration activity on event-based vision, denominated JPEG XE [19], it may be foreseen that a reference neuromorphic vision dataset may be available in a near future, as part of meaningful and precise test conditions and performance evaluation metrics to assess the performance of (new) NVDC solutions.

## V. NVDC: PERFORMANCE EVALUATION METRICS AND BENCHMARKING SOLUTIONS OVERVIEW

As mentioned in Section IV, the definition of precise performance evaluation metrics and benchmarking (anchor) solutions, together with meaningful test conditions, is of upmost importance for a proper NVDC solutions performance evaluation and comparison. In this context, this section goes one step further (with respect to Section IV) and provides an overview, in the form of a summary table (see TABLE 4), of the performance evaluation metrics and benchmarking coding solutions used to assess the performance of the currently available NVDC solutions (reviewed in Section III). Due to the extensive list of references, TABLE 4 also includes details on the Raw Data Type and Fidelity classification dimensions for better framing the performance evaluation metrics and benchmarking coding solutions. As in TABLE 2 and TABLE 3 (see Section IV), the NVDC references are presented in chronological order within each Raw Data Type class. In TABLE 4, '?' means that not enough information is provided in the reference to clarify the respective entry (as for the previous tables).

While the performance evaluation metrics are defined in TABLE 4, there are two that are common to all references, *CR* and *aggregation CR*; due to their importance, *CR* and *aggregation CR* are fully defined in the following. The *CR* corresponds to the ratio between the raw input data size in bits, where each uncompressed event has typically 64 bits long, i.e., 64 bits per event (bits/event), and the size in bits of the target event coding bitstream (aka *compressed bitstream size*); this CR is also known in the related literature as *End-to-End CR*, e.g., [25]. The *aggregation CR* is defined as the ratio between the size in bits of the event (temporal) aggregation based raw input data structure, e.g., 3D array of EFs, and the size in bits of the target event coding bitstream; the *aggregation CR* is also known in the related literature as *video encoder CR* [25] or as *Input-Output CR* [37]. Hence, the aggregation CR evaluates the NVDC solution performance with respect to the event data effectively coded. Recall that (raw input) data structuring based on event (temporal) aggregation may involve event discarding from the raw input data sequence (e.g., through event sampling [29]) and/or precision reduction of some event components (e.g., through timestamp component quantization and polarity accumulation [28]), as seen in Section III; all these operations induce information loss in the raw input data sequence, having, therefore, a direct impact on the amount of information to be coded.

From TABLE 4, the following conclusions can be derived:

- For both event and spike Raw Data Type classes, the CR is clearly the performance evaluation metric that has been more often adopted in the NVDC solutions performance assessment. However, for NVDC solutions involving event (temporal) aggregation based raw input data structures, it is common to find the aggregation CR as a CR replacement (although it can also be found as CR complement, e.g., in [25] and [37]). While a reference representation (or format) of the encoder (raw) input has been considered in all the CR computations within the same Raw Data Type class (e.g., in the event Raw Data Type class, the reference representation considers 64 bits/event), the same does not apply to the aggregation CR; the reference (i.e., uncompressed) representation considered in the aggregation CR varies with the type of event (temporal) aggregation performed (e.g., event count versus polarity accumulation). Considering aggregation CR based performance evaluation alone may make it difficult to compare NVDC solutions performance, due to the lack of a reference representation (format).
- Spatial distortion metrics, such as PSNR and SSIM of still intensity images reconstructed from decoded spike/event data, have been also adopted, to measure the loss induced by the NVDC solution, but only by

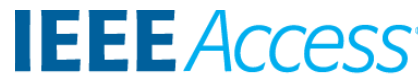

**TABLE 4.** Overview of the main performance evaluation metrics and benchmarking coding solutions used in currently available NVDC literature.

| Ref. | Raw Data Type | Fidelity | Main Performance Evaluation Metric(s) | Benchmarking Coding Solution(s) |
|---|---|---|---|---|
| Z. Bi *et al.*, 2018 [21] | Event | Lossless | - **CR** = $\frac{64 \times \text{Total \#events in event sequence [bits]}}{\text{Event coding bitstream size [bits]}}$ | - LZ77 [53];<br>- LZMA [54] |
| S. Dong *et al.*, 2019 [22] | Event | Lossless | - **CR** as in [21] | - [21];<br>- LZ77 [53];<br>- LZMA [54] |
| Y. Fu *et al.*, 2019 [24] | Event | Lossy | - **CR** as in [21] | - None |
| N. Khan *et al.*, 2021 [25] | Event | Lossy | - **CR** as in [21];<br>- **Aggregation (3D array of frames) CR** =<br>$\sum_{1}^{T_{frames}} \frac{\text{size of uncompressed frame [bits]}}{\text{size of compressed frame [bits]}}$<br>$T_{frames}$ = Total number of encoded frames | - [21];<br>- Several generic lossless data compression strategies adapted to NVDC |
| S. Banerjee *et al.*, 2021 [27] | Event | Lossy | - **CR** as in [21];<br>- **Event Distortion** = ***Spatial Distortion*** + ***Temporal Distortion***<br>*Spatial Distortion*: measured via **PSNR/SSIM** on frames obtained from event aggregation over timestamp bins.<br>*Temporal Distortion*: measured via proposed **temporal error metric** =<br>$T_{error} = \frac{1}{N_{fr}} \sum_{i=1}^{N_{fr}} \sqrt{\sum_{j} \left(T_{j,org} - T_{j,quant}\right)^2}$<br>$N_{fr}$ = # event volumes in a sequence<br>$T_{j,org}$ = Timestamp of j$^{th}$ event in the i$^{th}$ frame<br>$T_{j,quant}$ = Quantized timestamp of j$^{th}$ event in the compressed i$^{th}$ frame | - [22];<br>- [25] |
| I. Schiopu and R. C. Bilcu, 2022 [28] | Event | Lossy/Lossless | - For $\Delta = 10^{-6}$s: **CR** as in [21];<br>- For $\Delta$ = 5.555ms, 1ms and 0.1ms: **Aggregation (3D array of EFs) CR =**<br>$\frac{2 \times H \times W \times \text{Total \#EFs coded [bits]}}{\text{Event coding bitstream size [bits]}}$<br>$H$ = Event frame (EF) height in pixels<br>$W$ = EF width in pixels | - HEVC (4:0:0, Intra, lossless, FFmpeg implementation) [56];<br>- VVC (4:0:0, Intra, lossless) [57];<br>- CALIC (lossless) [58];<br>- FLIF (lossless) [59] |
| A. Hasssan *et al.*, 2022 [29] | Event | Lossy | - **CR** = $\frac{32}{(1-S) \times n}$<br>$S$ = Encoder output sparsity<br>$n$ = Pixel precision<br>32 = # bits used to represent an (raw) event ? | - None |
| I. Schiopu and R. C. Bilcu, 2022 [30] | Event | Lossy/Lossless | - **CR** as in [21];<br>- **Aggregation (3D array of EFs) CR** as in [28];<br>- **Average EF encoding speed [ms/EF]** = Ratio between the encoding time (ms) of the 3D array of event frames (EFs) and the number of EFs (in the 3D array of EFs) encoded | - HEVC (4:0:0, Intra, lossless, FFmpeg implementation) [56];<br>- VVC (4:0:0, Intra, lossless) [57];<br>- CALIC (lossless) [58];<br>- FLIF (lossless) [59] |
| M. Martini *et al.*, 2022 [31] | Event | Lossless | - **CR** as in [21] | - [21];<br>- LZMA [54] |

**TABLE 4.** ***(Continued.)*** **Overview of the main performance evaluation metrics and benchmarking coding solutions used in currently available NVDC literature.**

| | | | | |
|---|---|---|---|---|
| I. Schiopu and R. C. Bilcu, 2022 [32] | Event | Lossless | - **CR** as in [21];<br>- **Average event size [bits/event]** = Ratio between the compressed bitstream size in bits and the number of events in the raw input (asynchronous) event sequence;<br>- **Average encoded event rate [Mev/s]** = Ratio between the number of events in the raw input (asynchronous) event sequence and the encoding time;<br>- **Average event encoding/decoding speed [µs/event]** = Ratio between the encoding/decoding time (µs) of the raw input (asynchronous) event sequence and the number of events in the raw input (asynchronous) event sequence | - LZMA [54];<br>- ZLib [60];<br>- Bzip2 [61] |
| C. Wang *et al.*, 2023 [33] | Event | Lossless | - **CR** as in [21] | - [21];<br>- Zip [53] |
| I. Schiopu and R. C. Bilcu, 2023 [34] | Event | Lossy/Lossless | - **Aggregation (3D array of EFs) CR** as in [28];<br>- **Average EF encoding speed [ms/EF]** as in [30] | - [30];<br>- HEVC (4:0:0, Intra, lossless, FFmpeg implementation) [56];<br>- VVC (4:0:0, Intra, lossless) [57];<br>- CALIC (lossless) [58];<br>- FLIF (lossless) [59] |
| I. Schiopu and R. C. Bilcu, 2023 [35] | Event | Lossless | - **CR** as in [21];<br>- **Average event size [bits/event]** as in [32];<br>- **Average event encoding speed [µs/event]** as in [32];<br>- **Average encoded event rate [Mev/s]** as in [32] | - [32];<br>- LZMA [54];<br>- ZLib [60];<br>- Bzip2 [61] |
| B. Huang and T. Ebrahimi, 2023 [36] | Event | Lossless | - **CR** as in [21] | - [21];<br>- LZMA [54];<br>- Huffman coding [62];<br>- Sprintz delta [63] |
| S. Khaidem *et al.*, 2023 [37] | Event | Lossy | - **CR** as in [21];<br>- **Aggregation (3D array of frames) CR** as in [25] | - [21];<br>- [25];<br>- Several generic lossless data compression strategies adapted to NVDC |
| J. Adhuran *et al.*, 2024 [38] | Event | Lossy | - **CR** as in [21] | - [21];<br>- [25] |
| D. Stumpp *et al.*, 2024 [39] | Event | Lossy | - **CR** as in [21];<br>- **Event reduction [%]** = $100 \times \frac{N_S - N_{Tx}}{N_S}$<br>$N_S$ = # events in the input event sequence<br>$N_{Tx}$ = # events transmitted<br>- **Spatio-temporal event stream distance** = Asynchronous spatiotemporal spike metric (ASTSM) proposed in [64];<br>- **Temporal error [ms]** = Temporal distance of the events at a given pixel to the temporally closest event within a 3×3 window around the same pixel | - None |
| S. Dong *et al.*, 2019 [23] | Spike | Lossy | - **CR** with respect to raw spike data;<br>- **PSNR/SSIM** (from 1000 still intensity images reconstructed from raw and decoded spike data)<br>with *Distortion* = Intensity-based distance between spike trains =<br>$\lVert f_{s_1} - f_{s_2} \rVert = \sqrt{\frac{1}{K} \sum_{i=1}^{K} \left( \frac{1}{\Delta t_{s_1}^{(i)}} - \frac{1}{\Delta t_{s_2}^{(i)}} \right)^2}$<br>$K$ = # of Inter-Spike Intervals (ISIs)<br>$\Delta t_{s_1}^{(i)}$ = ISI sequence (converted from the spike train $f_{s_1}$ through $\bar{I} = \frac{\phi}{t}$ )<br>$\bar{I}$ = Average intensity of the pixel in the ISI period, t = ISI | - Raw spike data |

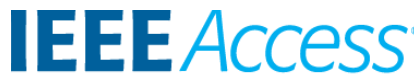

**TABLE 4.** ***(Continued.)*** **Overview of the main performance evaluation metrics and benchmarking coding solutions used in currently available NVDC literature.**

| | | | | |
|---|---|---|---|---|
| L. Zhu *et al.*, 2021 [26] | Spike | Lossy | - **CR** with respect to raw spike data;<br>- **PSNR/SSIM** (from 1000 still intensity images reconstructed from raw and decoded spike data)<br>with *Distortion* = Intensity-based distance between spike trains =<br>$\|f_{s_1} - f_{s_2}\| = \phi \sqrt{\frac{1}{K}\sum_{i=1}^{K}\left(\frac{1}{\Delta t_{s_1}^{(i)}} - \frac{1}{\Delta t_{s_2}^{(i)}}\right)^2}$<br>$\phi$ = Dispatch threshold<br>$K$ = # of Inter-Spike Intervals (ISIs)<br>$\Delta t_{s_1}^{(i)}$ = ISI sequence (converted from the spike train $f_{s_1}$ through $\bar{I} = \frac{\phi}{t}$)<br>$\bar{I}$ = Average intensity of the pixel in the ISI period, $t$ = ISI | - H.264/AVC [55] |

one event-based NVDC solution [27]; these spatial distortion metrics have been, however, adopted by all spike-based NVDC solutions [23], [26] in their performance assessment. Temporal error metrics have also been proposed for the same event-/spike-based NDVC solutions ([23], [26], and [27]) to measure the temporal distortion.

- Average event/EF encoding/decoding speed may also be found in the performance evaluation of some event-based NVDC solutions, notably the ones that target to be suitable for hardware implementation in low power/low cost event signal processing chips, e.g., [30], [32], [34], and [35]; in those solutions, it is also possible to find other event related performance assessment metrics such as the average event size and the average encoded event rate. It is important to note that all these metrics, except the average event size, are dependent on the hardware platform used which makes comparison rather difficult.
- In terms of benchmarking coding solutions, the situation seems a bit more controlled, in the sense that most of the NVDC solutions (within each Raw Data Type class) shares at least one benchmarking coding solution; still, the definition of a reference (anchor) benchmarking coding solution would be a step forward towards making the performance comparison task easier.

In summary, the main global conclusion that can be drawn from TABLE 4 is that the lack of precise common test conditions (and test material), among NVDC solutions sharing the same Raw Data Type classification, is making direct comparative performance analysis (based on experimental results) a difficult task. It is, therefore, urgent to define common test material and conditions (as well as benchmarking coding solutions) to promote solid and consistent advancements in the neuromorphic vision data coding area. As mentioned in Section IV, given the recent JPEG exploration activity on event-based vision (JPEG XE [19]), it may be foreseen that meaningful and precise test conditions and performance evaluation metrics (as well as anchor coding solutions) may be released in a near future to assess the performance of (new) NVDC solutions.

## VI. CONCLUSION

Neuromorphic vision data coding is currently a very hot research topic in multimedia representation and the recent JPEG exploration activity on event-based vision, denominated JPEG XE, is an acknowledgment of its practical importance. Over the last 6 years, several NVDC solutions have been developed, adopting rather different coding approaches, both for the lossless and lossy scenarios. In this challenging context, it is critical to understand the relationships between these multiple NVDC solutions in order the evolution of this technology can be faster and more solid. With this goal in mind, this paper proposes a classification taxonomy for NVDC solutions and reviews all the solutions available in the literature at the time this paper was written. This type of paper is essential to gather a systematic, high-level, and more abstract view of the NVDC landscape, thus allowing to better drive future research and standardization developments in this emerging technical area.

## VII. NVDC: CHALLENGES AHEAD

The literature overview presented in this paper shows that the NVDC area is still in its infancy. From this overview, notably from the conclusions drawn in sections II-C, IV, and V, it is possible to observe that, while some work has already been done, there are still several challenges associated with NVDC that need to be addressed, notably towards creating and developing a NVDC standard; this is, in fact, the main goal of the recent exploration activity on event-based vision launched by JPEG (JPEG XE [19]). Establishing a bridge to those conclusions, this section identifies some relevant challenges associated with NVDC, as follow:

- The lack of definition of meaningful and precise test conditions and performance evaluation methodologies, notably reflecting the NVD main applications and requirements, is a relevant challenge in the NVDC area (see sections IV and V). These test conditions and performance evaluation methodologies are essential to enable a proper performance comparison of NVDC solutions and, thus, to allow drawing the NVD coding *status quo* at any point in time, while fostering the development of increasingly efficient coding tools/solutions.

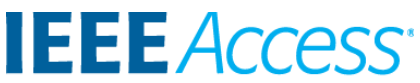

- Differently from conventional cameras, where all sensor pixels acquire visual information simultaneously at regular time intervals, in event cameras each sensor pixel asynchronously and independently triggers events driven by the visual scene dynamics, which leads to a variable data rate output. In fact, the data produced by event sensors may be sparse or somewhat dense depending on the frequency and location of lighting conditions changes and/or on the type of motion in the scene and/or on the event camera motion. Hence, for an event-based codec to be efficient, it must be able to accommodate different scene dynamics, that may result in sensor event sequences with rather different densities of events; standard-based point cloud coding solutions, e.g., [41], are a good example of the importance of considering the data sparsity in their development. Hence, this is certainly another relevant challenge in the NVDC area.
- Neuromorphic vision sensors made available by different manufacturers may have different sensitivities to the scene dynamics and produce event data sequences with different characteristics, e.g., in terms of spatial resolution and/or noise. Additionally, the current trend is for the new sensors to concurrently output event data and conventional intensity images or frames [65]. This may be a major challenge when it comes to the design and development of a NVDC solution: to be agnostic to the sensor characteristics, i.e., to have the same coding efficiency independently of the sensor model or characteristics.
- The current lack of a standardized NVD coding framework, essential to guarantee interoperability between different camera/sensor manufacturers and application frameworks, might be an obstacle to a rapid adoption of this type of cameras by the market and to the possible deployment of some applications where this type of visual information acquisition (sensing) may be profitable. The design and development of a standardized NVDC framework is, thus, another major challenge in the NVDC area, very much needed towards a fast and solid deployment of these technologies.
- The lack of visual subjective assessment methodologies and objective quality metrics for lossy coding is another challenge in the NVDC area. This challenge is particularly important when the output of the NVD-based applications is to be consumed by humans, e.g., HDR or high framerate video reconstruction. In this context, objective quality metric(s) independent of the task to be performed are essential to design and optimize future lossy event-based coding solutions, especially if the aim is to obtain a compressed representation suitable for several machine vision tasks and also for human consumption (after decoding).

Given the relevance and timeliness of the field, very exciting times are coming in the neuromorphic vision data coding arena.

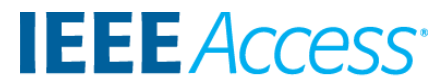

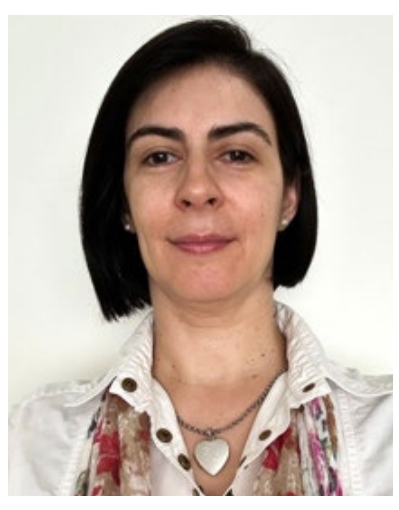

**CATARINA BRITES** (Member, IEEE) received the E.E., M.Sc., and Ph.D. degrees in electrical and computer engineering from the Instituto Superior Técnico, Universidade Técnica de Lisboa, Lisbon, Portugal, in 2003, 2005, and 2011, respectively.

She is currently an Assistant Professor with the Department of Information Science and Technology, Instituto Universitário de Lisboa (ISCTE), and a member of the Multimedia Signal Processing Group, Instituto de Telecomunicações, Lisbon. She has authored more than 70 international journals and conference papers. Her current research interests include 2D/3D video processing and coding, event data coding, plenoptic imaging representations, and machine learning. She serves as a technical program committee member for several international journals and conferences in the multimedia signal processing field. She is or has been an Associate Editor of IEEE Open Journal of Signal Processing and IEEE Transactions on Image Processing.

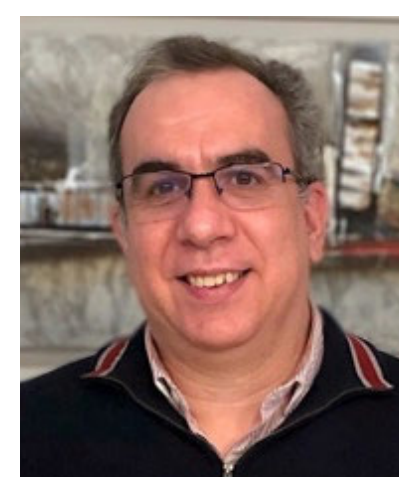

**JOÃO ASCENSO** (Senior Member, IEEE) received the E.E., M.Sc., and Ph.D. degrees in electrical and computer engineering from the Instituto Superior Técnico (IST), Universidade Técnica de Lisboa, Lisbon, Portugal, in 1999, 2003, and 2010, respectively.

He is currently an Associate Professor with the Department of Electrical and Computer Engineering, IST, and a member of the Instituto de Telecomunicações. He has authored more than 100 papers at international conferences. His current research interests include visual coding, quality assessment, light field and point cloud processing, and indexing and searching of multimedia content. He is an Associate Editor of IEEE Transactions on Image Processing and IEEE Transactions on Multimedia.